\documentclass[journal,12pt,onecolumn]{IEEEtran}
\IEEEoverridecommandlockouts
\usepackage{flushend}
\usepackage{balance}
\usepackage{cite}
\usepackage{amsmath,amssymb,amsfonts}
\usepackage{algorithmic}
\usepackage{graphicx}
\usepackage{textcomp}
\usepackage{xcolor}
\usepackage{algorithm}
\usepackage{comment}
\usepackage{tabu}
\usepackage{caption}
\usepackage{subcaption}
\usepackage{hyperref}
\usepackage{booktabs}
\usepackage[linesnumbered,lined,boxed,commentsnumbered,algo2e]{algorithm2e}
\def\BibTeX{{\rm B\kern-.05em{\sc i\kern-.025em b}\kern-.08em
    T\kern-.1667em\lower.7ex\hbox{E}\kern-.125emX}}

\ifCLASSOPTIONcompsoc
\else
\fi

\emergencystretch=\maxdimen
\usepackage{comment}
\usepackage{xcolor,colortbl}
\definecolor{green}{rgb}{0.1,0.1,0.1}
\newcommand{\CL}{\cellcolor{blue!50}}
\newcommand{\CLL}{\cellcolor{blue!25}}
\newcommand{\CLLL}{\rowcolor{gray!25}}

\newcommand{\blue}[1]{\textcolor[rgb]{0.00,0.00,0.00}{{#1}}}  
 \newcommand{\pink}[1]{\textcolor[rgb]{0.00,0.00,0.00}{{#1}}}

\begin{document}

\title{Machine Learning Testing in an ADAS Case Study Using Simulation-Integrated Bio-Inspired Search-Based Testing}

\author{\IEEEauthorblockN{Mahshid Helali Moghadam\IEEEauthorrefmark{1}\IEEEauthorrefmark{3},
Markus Borg\IEEEauthorrefmark{1},
Mehrdad Saadatmand\IEEEauthorrefmark{1}, 
Seyed Jalaleddin Mousavirad\IEEEauthorrefmark{2},
Markus Bohlin\IEEEauthorrefmark{3} and
Björn Lisper\IEEEauthorrefmark{3}}\\
\IEEEauthorblockA{\IEEEauthorrefmark{1}\small RISE Research Institutes of Sweden, Sweden}\\
\IEEEauthorblockA{\IEEEauthorrefmark{2}Universidade da Beira Interior, Covilhã, Portugal}\\
\IEEEauthorblockA{\IEEEauthorrefmark{3}Mälardalen University, Sweden}}

\maketitle

\begin{abstract}
This paper presents an extended version of {Deeper}, a search-based simulation-integrated test solution that generates failure-revealing test scenarios for testing a deep neural network-based lane-keeping system. In the newly proposed version, we utilize a new set of bio-inspired search algorithms, genetic algorithm (GA), $(\mu + \lambda)$ and $(\mu , \lambda)$ evolution strategies (ES), and particle swarm optimization (PSO), that leverage a quality population seed and domain-specific crossover and mutation operations tailored for the presentation model used for modeling the test scenarios. In order to demonstrate the capabilities of the new test generators within {Deeper}, we carry out an empirical evaluation and comparison with regard to the results of five participating tools in the cyber-physical systems testing competition at SBST 2021. Our evaluation shows the newly proposed test generators in {Deeper} not only represent a considerable improvement on the previous version but also prove to be effective and efficient in provoking a considerable number of diverse failure-revealing test scenarios for testing an ML-driven lane-keeping system. They can trigger several failures while promoting test scenario diversity, under a limited test time budget, high target failure severity, and strict speed limit constraints. 
\end{abstract}


\section{Introduction}
Machine Learning (ML) nowadays is used in a wide range of application areas such as automotive \cite{borg2019safely, borg2021exploring}, health care \cite{topol2019high} and manufacturing \cite{lee2018industrial}. Many of the ML-driven systems in these domains present a high level of autonomy \cite{cummings2020regulating} and meanwhile are subject to rigorous safety requirements \cite{burton2020mind}. In 2018, the European Commission (EC) published a strategy for trustworthy Artificial Intelligence (AI) systems \cite{EuropeanCommission}. In this strategy, AI systems are defined as ``systems that display intelligent behavior by analyzing their environment and taking actions--with some degree of autonomy--to achieve specific goals''. The EC states that a trustworthy AI system must be lawful, ethical, and robust. Self-driving cars are examples of safety-critical AI systems, which leverage various ML techniques such as Deep Neural Networks (DNN), machine vision, and sensor data fusion. Meanwhile, in the context of automotive software engineering, there is always a set of strict safety requirements to meet. 

The quality assurance methodology for AI systems \cite{hawkins2021AMLASguidance} is quite different from the conventional software systems, since the included ML components in those systems are not explicitly programmed, they are intended to learn from data and experience instead---called \textit{Software 2.0} \cite{Software2.0}. In addition, in AI systems, a part of the requirements is mainly seen as encoded implicitly in the data and the challenge of under-specificity is common in requirements definitions. However, it is still highly expected to assure the ability of the AI system to control the risk of hazardous events in particular in safety-critical domains. 

Self-driving cars, as one of the pioneering examples of safety-critical AI systems, are one of the focus points for research studies on quality assurance of ML-driven systems. \blue{There is also a vigorous need for the integration Verification and Validation (V\&V) of ML models that are deployed in self-driving cars to make sure that they are safe and dependable.} Many of the failures basically emerge in the interplay between software containing ML components, hardware, and remote sensing devices, e.g., sensors, cameras, RADAR, and LiDAR technologies. Hardware-In-the-Loop (HIL), simulation-based and field testing are common approaches for system-level verification of deployed ML models \cite{hawkins2021AMLASguidance}. System-level testing mainly targets defining a set of operational scenarios that could lead to failures. In this regard, in the ISO/PAS 21448 Safety of the Intended Function (SOTIF) standard \cite{SOTIF}---which addresses complementary aspects of functional safety in ISO 26262 \cite{ISO26262}---simulation-based testing has been considered a proficient approach and a proper complementary solution to the on-road testing. Testing on real-world roads is costly, does not scale to cover all the needed scenarios, and in addition, it is dangerous to create and execute critical scenarios. The use of virtual prototyping allows testing and verification at the early stages of the development and offers the possibility of efficient and effective testing. It can capture the whole of the operational environment to a great extent using high-fidelity autonomous driving simulators.  


\textbf{\textit{Research Challenge.}} In this study, we target an Advanced Driver-Assistance System (ADAS) that provides lane-keeping assistance. Effective and efficient system-level testing in simulation environments requires sophisticated approaches to generate critical test scenarios. The critical test scenarios are those that break or are close to break the safety requirements of the ADAS under test, which hence result in \textit{safety violations}. Generating effective test scenarios involves sampling from a large and complex set of test inputs. Several authors have shown the potential of search-based software test generation techniques to address this challenge. Various system-level testing techniques using different search-based testing approaches \cite{abdessalem2018learnable, gambi2019automatically, abdessalem2018testing, Moghadam2021Deeper, ebadi2021efficient, GABezier} for different types of ADAS, relying on simulators, have been proposed in recent years. 

\textbf{\textit{Research Contribution.}} 
In this paper, we present a bio-inspired computation-driven test generator, called {Deeper}, for effective and efficient generation of failure-revealing test scenarios to test a Deep Neural Network (DNN)-based lane-keeping system in the BeamNG driving simulator \cite{BeamNG}. The test subject is BeamNG.AI \footnote{\url{https://documentation.beamng.com/tutorials/ai/}.}, the built-in ML-driven driving agent in the BeamNG simulator. In this study, a failure is defined in terms of episodes in which the ego car---driven by the BeamNG.AI agent---drives partially outside the lane w.r.t a certain tolerance threshold. The tolerance threshold determines the percentage of the car’s bounding box needed to be outside the lane to be regarded as a failure.

{Deeper} in its current version in this paper benefits from the genetic algorithm (GA), $(\mu + \lambda)$ and $(\mu , \lambda)$ evolution strategies (ES) \cite{back1996evolutionary}, and the particle swarm optimization (PSO) \cite{kennedy1995particle} to generate failure-revealing test scenarios, which are test roads in our study. The problem is basically regarded as an optimization problem, and in order to generate the test scenarios that are of interest, we evaluate the quality of the test scenarios using a \textit{fitness (objective) function} that guides the search process to maximize the detected distance of the car from the center of the lane during driving of the car on the lane. The initial version of {Deeper}\cite{Moghadam2021Deeper} contained a test generator based on NSGA-II \cite{deb2002fast}. In this paper, we extend {Deeper} with four additional test generators based on GA, $(\mu + \lambda)$ and $(\mu , \lambda)$ ESs, and PSO. In the newly proposed test generators, we leverage an initial quality population seed to boost the search process, and also customize common crossover and mutation operations for the presentation model used for the test scenarios in the search algorithms---develop crossover and mutation operations tailored for model. We rely on the presentation model \cite{riccio2020model} used by DeepJanus \cite{riccio2020model} based on Catmull-Rom cubic splines \cite{catmull1974class}.  
      
\textit{Empirical evaluation.} In order to carry out an empirical evaluation, we use the setup provided by the cyber-physical systems (CPS) testing competition\footnote{\url{https://github.com/se2p/tool-competition-av}.} at the IEEE/ACM 14\textsuperscript{th} International Workshop on Search-Based Software Testing (SBST). Our experiments are designed to answer three main research questions which are as follows:\\
\textbf{RQ1:} How capable are these test generators to trigger failures?\\
\textbf{RQ2:} How diverse are the generated failure-revealing test scenarios?\\
\textbf{RQ3:} How effectively and efficiently do the test generators perform? \blue{It has three parts, which are as follows: Given a certain test budget, 3.1) how many test scenarios are generated, 3.2) what proportion of the scenarios is valid, and finally, 3.3) what proportion of the valid test scenarios leads to triggering failures?} \\
We provide a comparative analysis on the performance of the proposed bio-inspired test generators in {Deeper} and five counterpart tools all integrated into the BeamNG simulator. In this regard, we compare the results of the proposed test generators in {Deeper} with five other test generator tools, Frenetic \cite{Frenetic}, GABExploit and GABExplore \cite{GABezier}, Swat \cite{SWAT}, and also the earlier version of the {Deeper} (based on NSGA-II) \cite{Moghadam2021Deeper}---all participating tools in the CPS competition at SBST 2021. In order to do a fair comparison, we consider the same experimental evaluation procedures as the original CPS tool competition. 
Our experimental results show that first, the newly proposed test generators in {Deeper} present a considerable improvement on the previous version, second, they perform as effective and efficient test generators that can provoke a considerable number of diverse failure-revealing test scenarios w.r.t different target failure severity (i.e., in terms of tolerance threshold), available test budget, and driving style constraints (e.g, setting speed limits). For instance, in terms of the number of triggered failures within a given test time budget and with less strict driving constraints, the $(\mu + \lambda)$ ES-driven test generator in {Deeper} considerably outperforms other tools while keeping the level of promoted failure diversity quite comparable to the counterpart tool with the highest number of detected failures in the competition. Then, with respect to the test effectiveness and efficiency in non-strict conditions, again {Deeper} $(\mu + \lambda)$ ES-driven test generator results in the highest effectiveness in terms of the ratio of the number of detected failures to the generated valid test scenarios, while PSO-, and GA-driven are the next effective ones, which show performance comparable to the top counterparts.    
Meanwhile, as a distinctive feature, none of the newly proposed test generators leaves the experiment without triggering any failures, and in particular, they act as more reliable test generators than most of the other tools for provoking diverse failures under a limited test budget and strict constraints.


The rest of this paper is organized as follows: \blue{Section \ref{D:Sec::background} presents background information on bio-inspired search techniques including evolutionary and swarm intelligence techniques, and simulation-based ADAS testing.} Section \ref{D:Sec::Deeper} presents the problem formulation and the technical details of our proposed test generators in {Deeper}. Section~\ref{D:sec::EmpiricalEvaluation} elaborates on the empirical evaluation, including the research method and experiments setup. Section \ref{D:Sec::Result&Discussion} discusses the results, answers to the RQs, and the threats to the validity of the results. Section \ref{D:Sec::RelatedWork} provides an overview of the related work, and finally, Section \ref{D:sec::summary} concludes the paper with our findings and the potential research directions for future work.

\section{Background} \label{D:Sec::background}
\subsection{Evolutionary and swarm intelligence} These algorithms are two main classes of random search techniques, which are widely used in many different optimization problems. Genetic algorithms (GA) and evolution strategies (ES) are two of the the main categories inside the family of evolutionary algorithms (EAs). Particle swarm optimization (PSO) is one of the primary representatives of swarm intelligence algorithms \cite{slowik2020evolutionary}.

\textbf{Genetic algorithms} is one of the most common nature-inspired optimization techniques. It starts with a random population of individuals---each called a chromosome---representing a potential solution for the problem. The objectives to be optimized in the problem are defined in an \textit{objective function} and the quality of the solutions is measured via this function. It shows how "well'' each solution satisfies the objective. The quality of each individual, which is also referred to as ``fitness'', is a main factor during the evolution process. At each generation, a new population is formed based on the selected individuals from the previous generation. Three operations are involved in forming the new generation, which are as follows:
\begin{enumerate}
    \item Selection, which mainly identifies highly-valued individuals from the previous generation.
    \item Crossover, which breeds ``child'' individuals by exchanging parts of the ``parent'' individuals. The child individuals (offspring) are formed by selecting genes from each parent individual.
    \item Mutation, which applies small random adjustments to the individuals.   
\end{enumerate}

Crossover and mutation operations are applied w.r.t user-set probabilities, and these two operations might be used, either independently or jointly, to create new individuals to form a new population. The resulting individuals are added to the new population. The fitness values are calculated and stored for each individual in this population. This process iterates each generation until stopping criteria are met, e.g., a user-set number of generations or an allowed time budget is exhausted \cite{ebadi2021efficient, slowik2020evolutionary}.

\textbf{Evolution strategy} is another common class of EAs. It is commonly used in almost all fields of optimization problems including discrete and continuous input spaces. ES also involves applying selection, recombination, and mutation to a population of individuals over various generations to get iteratively evolved solutions. Two canonical versions of ES are  $(\mu/\rho + \lambda)$ and $(\mu/\rho ,\lambda)$ evolution strategies. If $\rho=1$, we have ES cases without recombination, which are denoted by  $(\mu + \lambda)$ and $(\mu,\lambda)$ ESs. In the case of $\rho=1$, the recombination is simply making a copy of the parent. In these notations, $\lambda$ and $\mu$ indicate the size of the offspring and population respectively. The main difference between GA and ES is related to the selection step. In GA, at each iteration, the next generation is formed by selecting highly-valued individuals, while the size of the population is kept fixed \cite{slowik2020evolutionary}. In ES, a temporary population with the size of $\lambda$ is created and the individuals in this temporary population undergo crossover and mutation at user-set probabilities regardless of their fitness values. In $(\mu + \lambda)$ ES, then, both parents and the generated offspring resulting from the temporary population are copied to a selection pool---with size $(\mu + \lambda)$---and a new population with size $\mu$ is formed by selecting the best individuals. While, in $(\mu,\lambda)$ ES, the new generation with size $\mu$ is selected only from the offspring (with size $\lambda$). Therefore, a convergence condition as $\mu < \lambda$ is required to guarantee an optimal solution \cite{beyer2002evolution}.

\textbf{Particle swarm optimization} is one of the most common representative of the swarm intelligence (SI) algorithms, which form a big class of nature-inspired optimization methods alongside the evolutionary algorithms. The SI algorithms present the concept of collective intelligence, which is mainly defined as a  collective behavior in a group of individuals that seem intelligent. SI algorithms have been inspired from collective behavior and self-organizing interactions between living agents in the nature, e.g., ant colonies and honey bees
\cite{mavrovouniotis2017survey}.

PSO is an optimization method simulating the collective behavior of certain types of living species. In PSO, \textit{cooperation} is an important feature of the system as each of the individuals changes its searching pattern based on its own and others' experiences. PSO starts with a swarm of random particles. Each particle has position and velocity vectors, which are updated w.r.t the local and global best values. The best values get updated at each iteration. In the application of PSO, each particle (individual) represents a potential solution and is often modeled as a vector containing $n$ elements, in which each element represents a variable of the problem that is being optimized. Like GA, PSO searches for the optimal solution through updating solutions and creating subsequent generations, though without using evolution operators \cite{slowik2017nature}. The position (the elements) and velocity of each particle are updated as follows:
\begin{equation} \label{eq:position_update}
P^{t+1} = P^{t} + V^{t+1} 
\end{equation}

\begin{equation} \label{eq:Velocity_update}
V^{t+1} = wV^{t} + c_1 r_1 (P^{t}_{best}- P^{t}) + c_2 r_2 (G^{t}_{best} -P^{t})
\end{equation}
where $P^{t}$ and $V^{t}$ are the position and velocity of the particle at iteration $t$, respectively; $P^{t}_{best}$ and $G^{t}_{best}$ are the local best position of the particle and the global best one up to the iteration $t$. The first part of (\ref{eq:Velocity_update}) is perceived as \textit{inertia}, which indicates the tendency of the particle to keep moving in the same direction, while the second part---which reflects a cognitive behavior---indicates the tendency towards the local best position discovered by the particle and the last part---which is the social knowledge---reflects the tendency to follow the best position found so far by other particles.

Therefore, at each iteration, the position of each particle is updated based on its velocity, and the velocity is controlled by the inertia and accelerated stochastically towards the local and global best values. \blue{$r_1$ and $r_2$ are random weights from range $[0, 1]$ which adjust the cognitive and social acceleration.} In (\ref{eq:Velocity_update}), $w$ is inertia weight, which adjusts the ability of the swarm to change the direction and makes a balance between the level of exploration and exploitation in the search process. A lower $w$ leads to more exploitation of the best solutions found, while a higher value of $w$ facilitates more exploration around the found solutions. $c_1$ and $c_2$ are the acceleration hyperparameters defining to what extent the solutions are influenced by the local best solutions and global best solution. These hyperparameters and the inertia weight could be static or changed dynamically over the iterations. For instance, $w = 0.72984$ and $c1+c2>4$ is a common setup for a static configuration of these parameters \cite{clerc2002particle}.

\subsection{Simulation-based ADAS Testing} 
\begin{figure}[b]
  \centering
  \includegraphics[width=.80\columnwidth, height= 7cm]{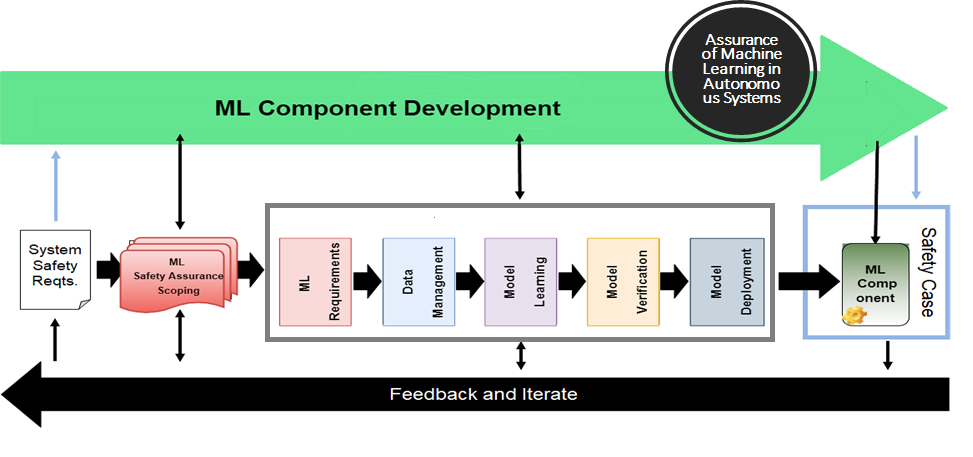}
  \caption{\blue{A general view of AMLAS process, adapted from \cite{hawkins2021AMLASguidance}.}}
  \label{fig:AMLAS}
\end{figure}

\blue{System-level testing for systems with ML-driven components in safety-critical domains such as automotive industry introduces new challenges for system quality certification and verification. In this regard, new safety complementary standards and guides are always developed to cover the gaps and ensure the safety of ML components in those domains. Standards such as ISO 26262 Functional Safety \cite{ISO26262} and ISO 21448 Safety of the Intended Functionality \cite{SOTIF} often provide a high-level view of the requirements that are required to be satisfied in a safety case for an ML-driven system. However, how it shall be implemented and argued for being adequate and sufficient is highly up to the specific stakeholder organization \cite{borg2022ergo}. In this regard, Hawkins et al. \cite{hawkins2021AMLASguidance} introduced a methodology for the Assurance of ML in Autonomous Systems, called AMLAS. It presents a systematic process for integrating safety assurance into the development of ML systems and introduces verification procedures at different stages, e.g., learning model verification and system-level (integration) verification that happens after integrating the ML model into the system. Figure \ref{fig:AMLAS} provides an overview of the proposed AMLAS framework.}  

\blue{Assurance activities on ML safety requirements, data management, model learning alongside the model verification and model deployment \cite{hawkins2021AMLASguidance, riccio2020testing} are different levels  proposed by AMLAS. In this regard, model verification/testing and model deployment assurance---which is typically satisfied by integration and system-level testing---are considered different test levels for the quality assurance of ML systems. Various testing or formal verification techniques could be used to assure that the ML model satisfies the ML requirements. It is important to note that the ML testing is performed on the verification data set that is not involved and used in the training process in anyhow \cite{hawkins2021AMLASguidance}. Model deployment assurance involves assuring the satisfaction of system safety requirements after the integration of the model into the system w.r.t the specified Operational Design Domain (ODD) for the system. Model testing could be regarded as unit testing for ML components and integration testing focuses on the issues emerging after the integration of the ML model into the system.
Another view on testing ML systems is based on access to the test subject. In this regard, there are black-box and white-box testing approaches analogous to traditional non-ML systems. Black-box testing involves access only to the ML inputs and outputs, while white-box testing implies access to the internal architecture of the test ML subject, code, hyperparameters, and training/test data. In this regard, Riccio et al. \cite{riccio2020testing} introduced another type of ML testing called data-box, which requires access to data plus everything that a black-box test requires.} 

\blue{The use of virtual testbeds, e.g., Simulation-based testing, has been considered a feasible, non-expensive, effective, and proper complement to field testing \cite{scharke2022virtual}. In addition to the possibility of early-stage verification, virtual testing provides the possibility to create a huge number of test scenarios and examine autonomous driving functions in critical and unsafe situations. Moreover, the test could be reproducible and scalable. Recently, a growing number of commercial and open-source simulators have been developed to support the need for realistic simulation of self-driving cars---we refer interested readers to a review by Rosique et al.~\cite{rosique2019systematic}. In this regard, for instance, 
various high-fidelity simulators such as the ones using physics-based models (e.g., SVL simulator \cite{SVLSimulator}, Pro-SiVIC \cite{ProSivic_belbachir2012simulation}, and PreScan \cite{PreScan}) and the ones based on game engines (e.g., BeamNG.tech \cite{BeamNG} and CARLA \cite{dosovitskiy2017carla}) have considerably contributed to this area by providing the possibility of realistic simulations of functionalities in autonomous driving. Kaur et al. \cite{kaur2021survey} also provide a comparison of some of the common simulators w.r.t some essential features expected from the autonomous driving simulators such as perception, multi-view geometry, path planning, 3D virtual environment, traffic infrastructure and some other relevant features that can be crucial to assure the fidelity of the virtual environment to reality.}

\section{{Deeper}: A Bio-Inspired Simulation-Integrated Testing Framework}\label{D:Sec::Deeper}
This section presents the technical details of {Deeper} and shows how it challenges a DNN-based lane-keeping system in a self-driving car trained and tested in the BeamNG simulator environment \cite{BeamNG}. The subject system is a built-in AI driving agent encompassing a steering angle predictor (ML model) which receives images captured by an onboard camera in the simulation environment. Then, the test inputs (test cases) generated by {Deeper} are defined as scenarios in which the car drives. Our target is to generate diverse test scenarios triggering the misbehavior of the subject system. In this regard, we benefit from bio-inspired search-based techniques to explore the input space and generate highly-valued failure-revealing test scenarios. 

\subsection{Test Scenario and Failure Specification}\label{D:Sec::Test_Scenario_Spec}
\begin{figure}
  \centering
  \includegraphics[width=.55\columnwidth, height = 4cm]{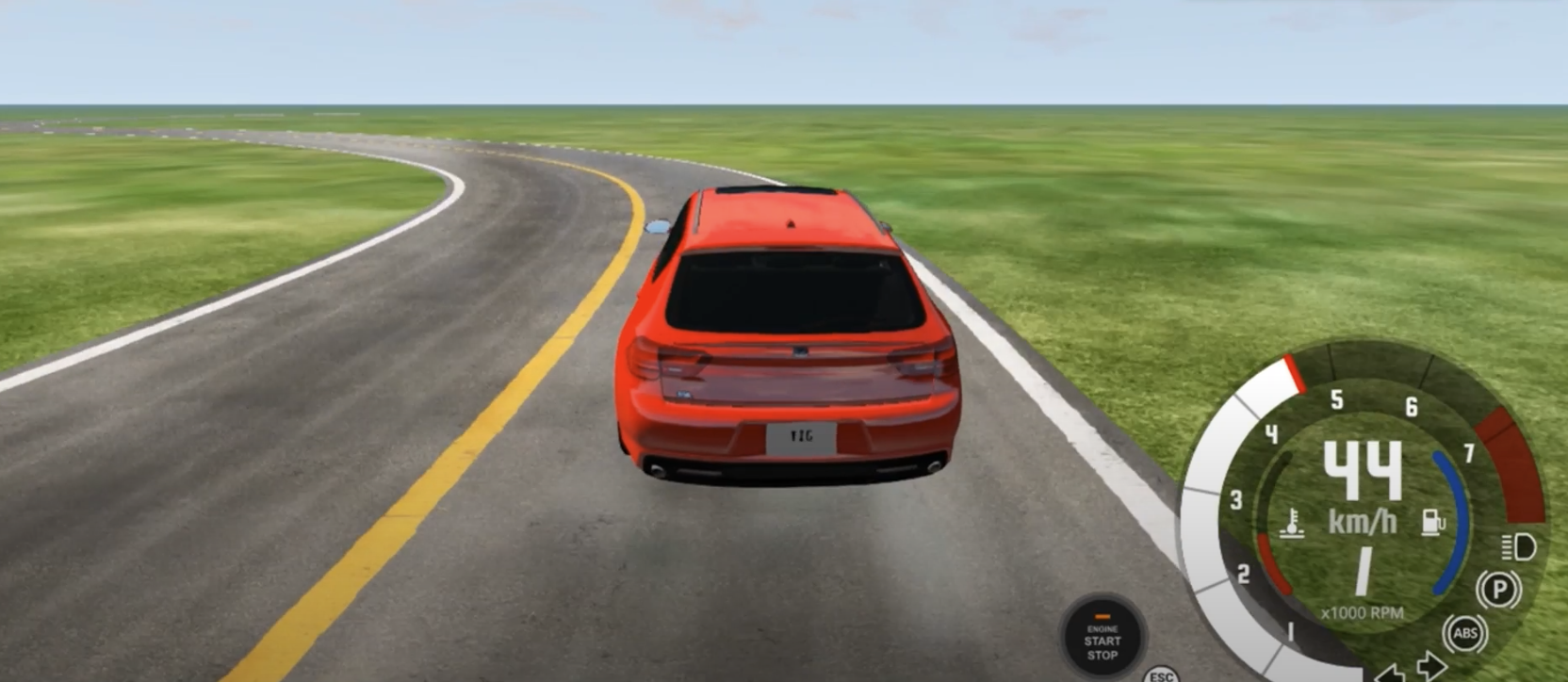}
  \caption{A test scenario executed in the BeamNG simulator.}
  \label{fig:TestScenarioSample}
\end{figure}

Test scenarios are defined as combinations of roads, the environment including e.g., the weather and illumination, and the driving path, i.e., starting and end points and the lane to keep. Hereafter, we consider scenarios involving a single asphalt road surrounded by green grass where the car is to drive on the right lane, and the environment is set to a clear day with the overhead sun (see Figure \ref{fig:TestScenarioSample}). Therefore, the focus of {Deeper} is to generate diverse roads which trigger failures in the system under test. In this system, failure is defined in terms of an episode, in which the car drives partially outside the lane meaning that $X\%$ of the car’s bounding box gets outside the lane. $X$ is a configurable tolerance threshold for {Deeper}.

In the test scenarios, the road composes two fixed-width lanes with a yellow center line and two white lines separating the lanes from the non-drivable area. In BeamNG, each road is mainly described by a set of points that are used by the simulation engine to render the road. The simulation engine accomplishes the rendering by interpolating the points and creating a sequence of polygons on the points---provided by the SBST competition setup \footnote{\url{https://github.com/se2p/tool-competition-av.git}.}. Notably, not every sequence of road points results in valid roads, so each sequence of points is also validated against some initial geometrical constraints related to the road polygons and some other domain-specific constraints. The main constraints are: 1) the start and end points of the road shall be different, 2) the road shall be completely contained in the map used in the simulation, 3) the road shall not self-intersect, and 4) the road shall not contain too sharp turns that force the vehicle to invade the opposite lane. To assure the satisfaction of these constraints, {Deeper} validates the generated roads before getting executed and consequently, the invalid roads are not counted as failed test scenarios.          

\begin{figure}
  \centering
 \begin{subfigure}[b]{0.32\textwidth}
         \centering
         \includegraphics[width=\textwidth, height =5cm]{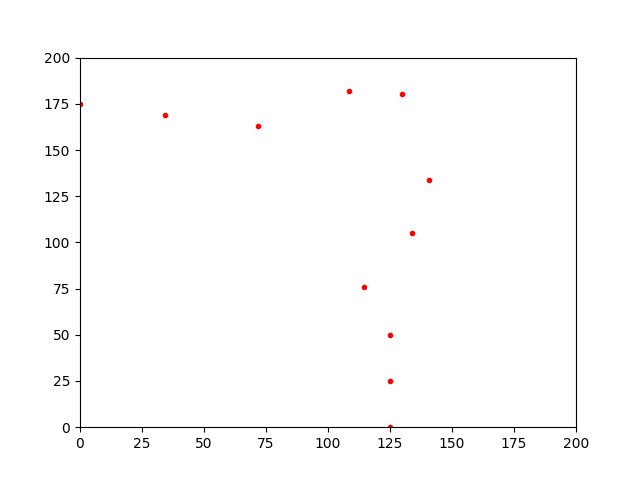}
         \caption{Control Points}
         \label{fig:CP}
     \end{subfigure}
     \begin{subfigure}[b]{0.32\textwidth}
         \centering
         \includegraphics[width=\textwidth,  height =5cm]{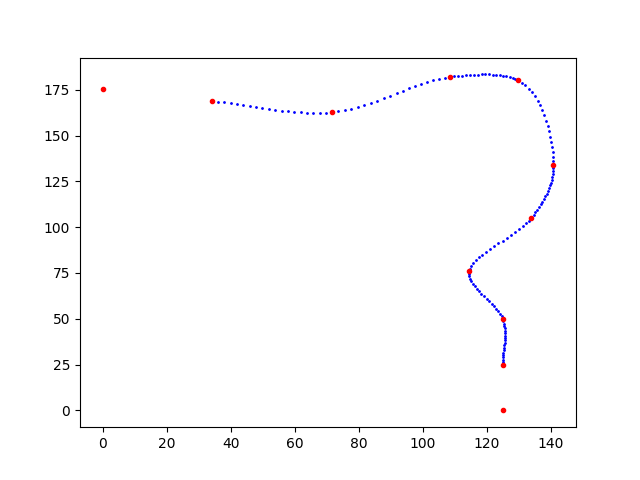}
         \caption{Calculated Sample Points }
         \label{fig:SP}
     \end{subfigure}
     \begin{subfigure}[b]{0.3\textwidth}
         \centering
         \includegraphics[width=\textwidth,  height =4cm]{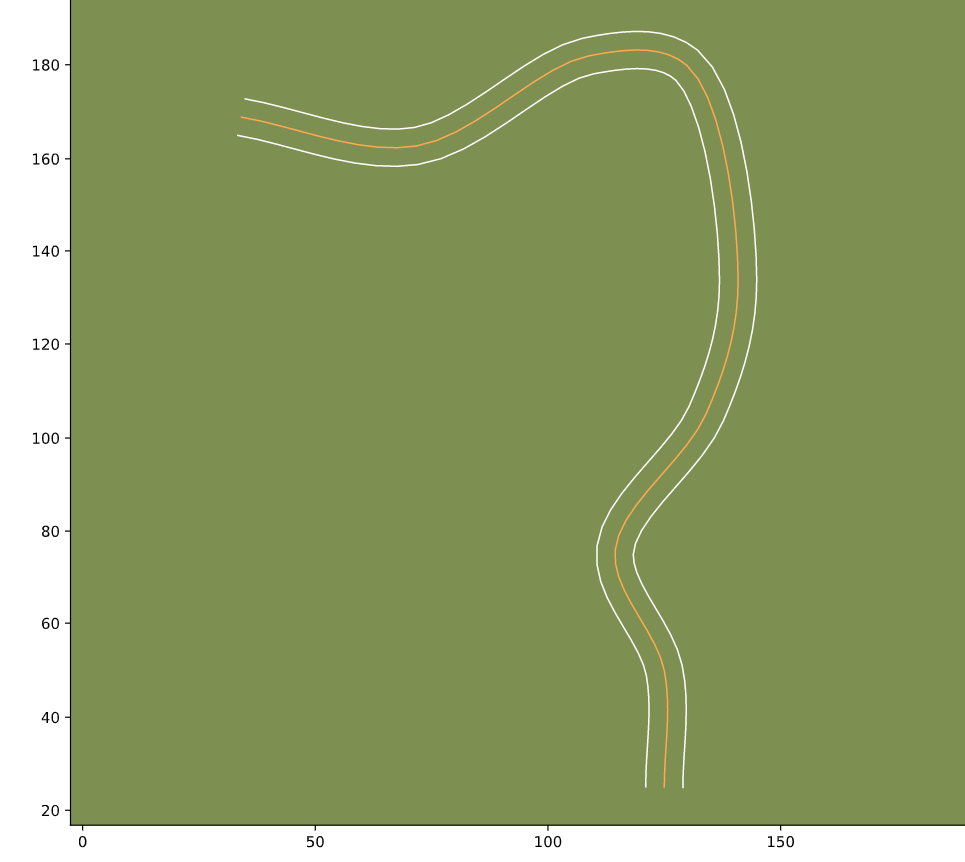}
         \caption{The corresponding road in the simulation}
         \label{fig:RenderedRoad}
     \end{subfigure}
        \caption{The representation model of a road}
        \label{fig:ModeloftheRoad}
\end{figure}

\textit{Road Representation Model:} In order to convert the abstract road model into a proper set of points that can be rendered by the simulation engine, we rely on the representation model used by {DeepJanus} \cite{riccio2020model} based on Catmull-Rom cubic splines \cite{catmull1974class}. \blue{The Catmull-Rom splines is a method that involves defining a series of points at intervals and a piecewise-defined function to calculate additional points within the intervals. The constructed curve will pass through the given set of points and feature a certain number of continuous derivatives. In Catmull-Rom interpolating splines, those given points defining the spline are called "control points".} Therefore, each road is represented by two sets of points, \textit{control points} and \textit{sample points}. First, control points are provided as input for a candidate road. Second, sample points are calculated using the Catmull-Rom calculation algorithm. Third, the simulation engine uses the sample points, if they are valid, to render the road. Figure \ref{fig:ModeloftheRoad} shows the representation model of a road in terms of control and sample points (\ref{fig:CP} and \ref{fig:SP}) and the corresponding rendered road in the simulation (\ref{fig:RenderedRoad}). 

\begin{eqnarray}
CP= \langle C_1, C_2, \cdots, C_m \rangle  \ , \  &{R_i}_{min} \leq C_i \leq {R_i}_{max}\\
&{R_i}_{min}, \ {R_i}_{max} \in R
\end{eqnarray}

\begin{eqnarray}
SP= \langle S_1, S_2, \cdots, S_n \rangle  \ , \  &SP = Catmull\_Rom\_Spline (CP)\\
&{R_i}_{min} \leq S_i \leq {R_i}_{max}\\
&{R_i}_{min}, \ {R_i}_{max} \in R
\end{eqnarray}

\subsection{Fitness Function}\label{D:Sec::Fitness_Function}
As indicated above, the focus is the generation of diverse failure-revealing test scenarios w.r.t the intended tolerance threshold. The competition setup elaborates the positional data and detects the episodes in which the car breaks out of the lane bounds, i.e., out of bound episodes (OBE). It computes the distance of the car from the center of the lane in those OBE episodes and reports an OBE failure \cite{gambi2019asfault} each time that the car drives outside the lane if the percentage of the area of the car that is outside the lane is bigger than the intended threshold (referred to as $X$ in Section~\ref{D:Sec::Test_Scenario_Spec}).    

The problem is regarded as an optimization problem, and in order to generate valuable test scenarios leading us to meet the target, we evaluate the quality of the test scenarios using a \textit{fitness (objective) function}. In this regard, for each test scenario, the main objective of interest to be optimized is the maximum detected distance of the car from the center of the lane during driving of the car on the lane. So, more accurately the fitness function that we want to minimize is as follows:

\begin{equation}
   Fitness\_Function = (lane\_width)/2 - d(center\_polyline, car\_position) 
\end{equation}
where $lane\_width$ is the width of the lane and $d(spin\_polyline, car\_position)$ indicates the distance of the car position from the central polyline (center) of the lane.  

\subsection{Bio-Inspired Search Algorithms}\label{D:Sec::Search_Algorithms}
We are interested in sampling from the space of possible test scenarios in an effective way to generate those that lead to the emergence of failures. This can be achieved by using an optimization algorithm to guide the search by the fitness function. Therefore, in order to find the solutions of interest, we use evolutionary and swarm intelligence algorithms i.e., GA, $(\mu + \lambda)$ and $(\mu , \lambda)$ ESs, and PSO, guided by the fitness function introduced in Sec.\ref{D:Sec::Fitness_Function}.
\blue{In {Deeper} the roads---as the main focus of the test scenarios---are evolved through changes applied to the control points by the bio-inspired algorithms. 
Any changes in the set of control points does not lead to creating a valid road. Once a new set of control points is formed, the simulation engine performs the road rendering by first interpolating the points, then checking the satisfaction of the geometrical and domain-specific conditions.} 

The search algorithms are mainly modeled on the basis of population evolution over time and they usually get started with the creation of a random population of solutions. In {Deeper}, we leverage an initial quality population seed to boost the search process regarding the fact that the search is done at a fixed test budget. Throughout the development, the impact of different initial population seeds was investigated. For instance, starting from an initial random population seed was not quite effective to lead the search to find the failure-revealing solutions w.r.t high tolerance thresholds within a reasonable test budget. \blue{We created a quality population seed mainly based on the use of an improved random road generation technique, which besides tweaking the control points, changes the geometrical parameters of the roads such as the maximum allowed angle and the length of road segments as well. We formed 90\% of the quality population seed through the generated set of roads from the improved random generation technique. Then, in order to augment the population seed, we filled the remaining 10\% of the population with solutions randomly extracted from a prior standalone 5-hour execution of Deeper NSGA-II, which had started from a random population seed---generated through just tuning the control points. For this latter part, the generated solutions which could cause OBEs w.r.t a tolerance threshold $\tau \geq 0.5$ were only considered. We use this quality population seed for running Deeper NSGA-II in the empirical evaluation as well. Basically, the same population seed is used for running all the test generators involved in Deeper.}


\subsubsection{\textbf{Genetic Algorithm}} \label{D:Sec::Search_Algorithms_GA} 
The GA-driven test generator in {Deeper} starts with forming an initial population by sampling from the quality population seed. Over various generations, new populations of test scenarios are formed through applying crossover and mutation operations to the best ones selected from the previous generations.

\textit{Selection:} We use \textit{tournament selection} for identifying the promising test scenarios, which have a high probability to lead to failures and safety violations. In tournament selection, a subset of the population is selected at random in each tournament and the best test scenario of the subset is picked. The number of individuals participating in each tournament indicates the size of the tournament. 

\begin{figure}[h]
  \centering
   \begin{subfigure}[b]{0.3\textwidth}
         \centering
         \includegraphics[width=\textwidth, height =5cm]{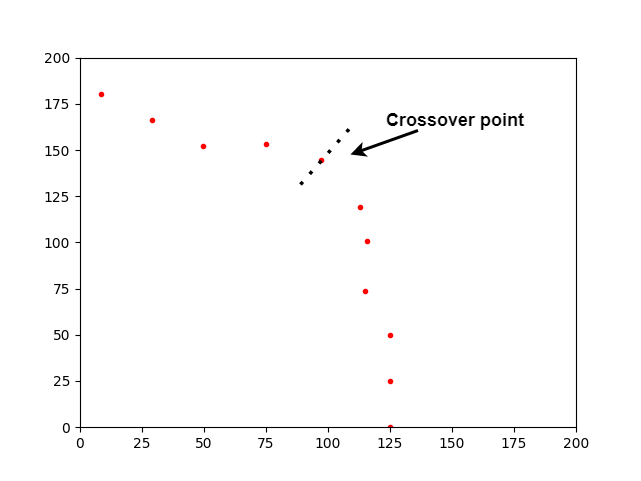}
         \caption{Parent Road 1: Control Points}
         \label{fig:Road1}
     \end{subfigure}
     \begin{subfigure}[b]{0.3\textwidth}
         \centering
         \includegraphics[width=\textwidth,  height =5cm]{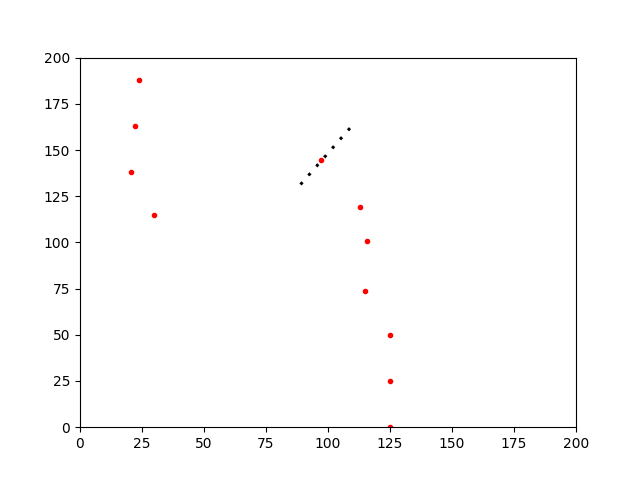}
         \caption{Child Road 1: Control Points}
         \label{fig:Road3_Control_points}
     \end{subfigure}
     \begin{subfigure}[b]{0.3\textwidth}
         \centering
         \includegraphics[width=\textwidth,  height =5cm]{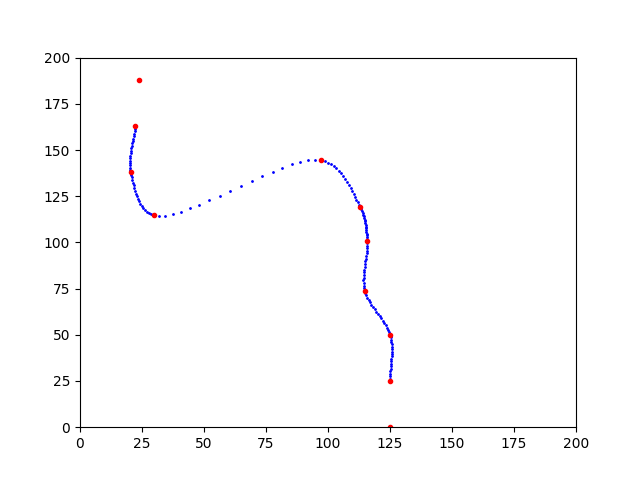}
         \caption{Child Road 1: Sample Points}
         \label{fig:Road3_Sample_points}
     \end{subfigure}
     \newline
     \begin{subfigure}[b]{0.3\textwidth}
         \centering
         \includegraphics[width=\textwidth, height =5cm]{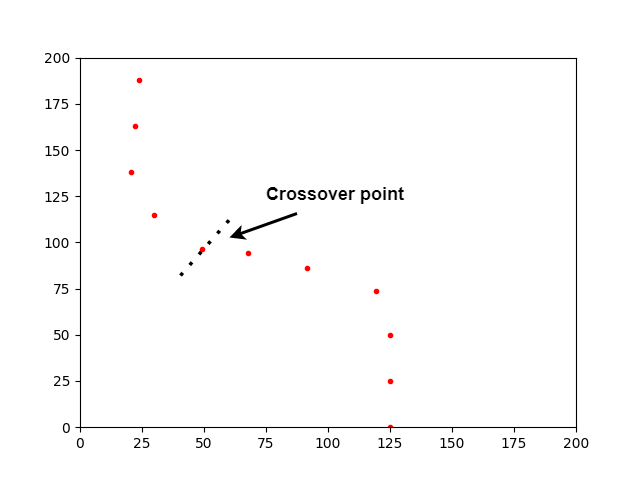}
         \caption{Parent Road 2: Control Points}
         \label{fig:Road2}
     \end{subfigure}
     \begin{subfigure}[b]{0.3\textwidth}
         \centering
         \includegraphics[width=\textwidth,  height =5cm]{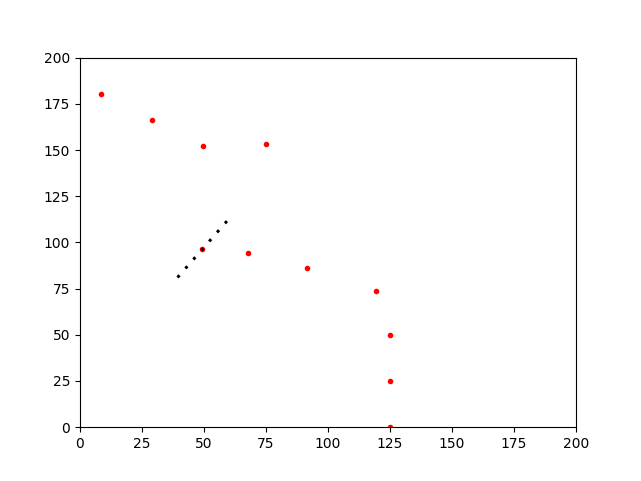}
         \caption{Child Road 2: Control Points}
         \label{fig:Road4_Control_points}
     \end{subfigure}
     \begin{subfigure}[b]{0.3\textwidth}
         \centering
         \includegraphics[width=\textwidth,  height =5cm]{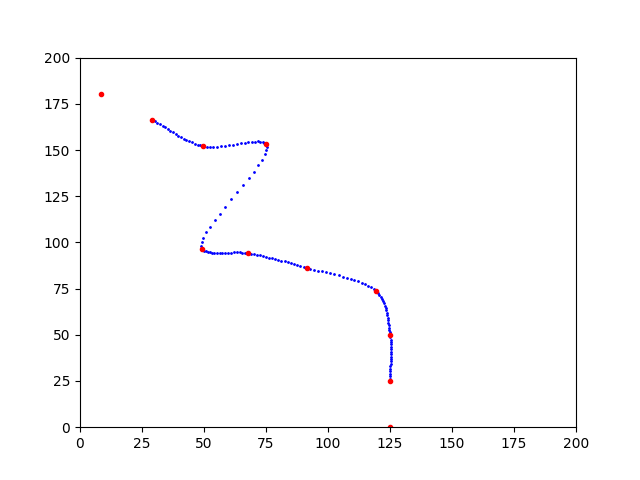}
         \caption{Child Road 2: Sample Points}
         \label{fig:Road4_Sample_points}
     \end{subfigure}
     \caption{Crossover operation on two sample roads}
      \label{fig:Crossover}
\end{figure}

\textit{Crossover:} We develop a domain-specific \textit{one-point crossover} operation tailor-made to our representation model.
The proposed crossover operation performs the segment exchange at the level of \textit{control points}, which means that a random point is selected as the crossover point in the sets of control points in the parent roads, then the parts of the sets beyond the crossover point are swapped between the parents, and accordingly, two new sets of control points for two child roads are formed. The corresponding sample points for the generated child roads are calculated using the Catmull-Rom calculation algorithm (see Figure \ref{fig:Crossover}).
However, still, these resulting sets of points might not represent valid roads w.r.t the geometrical constraints, so before adding these new test scenarios to the offspring, we also check their validity and let only the resulting valid roads be added to the offspring. If both generated child roads are valid, then both of them will be transferred to the offspring, while if one of them is valid, we keep the valid child and another crossover point is tried to breed the parent roads and generate the second valid child. Likewise, if none of the child roads are identified as valid roads, another crossover point is tried. All in all, in order to generate two valid child roads from a crossover operation, at most five attempts to try with different crossover points are done. If in the end, the attempts do not lead to valid child roads, the whole process will be rolled back and the original parents will be added to the offspring. 

\begin{figure}[h]
  \centering
 \begin{subfigure}[b]{0.35\textwidth}
         \centering
         \includegraphics[width=\textwidth, height =5cm]{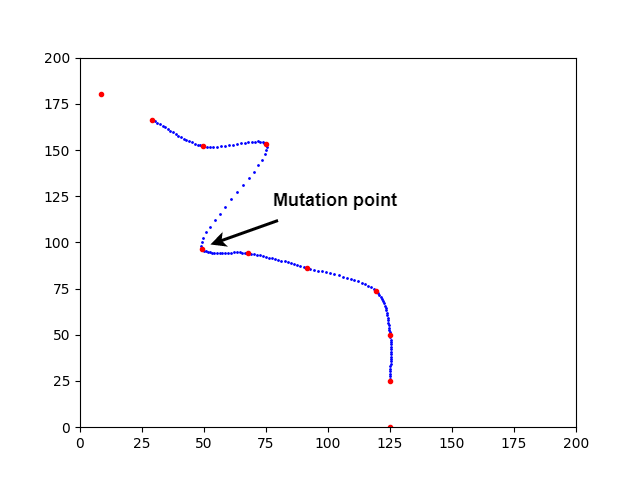}
         \caption{Original road}
         \label{fig:Original_Road}
     \end{subfigure}
     \begin{subfigure}[b]{0.35\textwidth}
         \centering
         \includegraphics[width=\textwidth,  height =5cm]{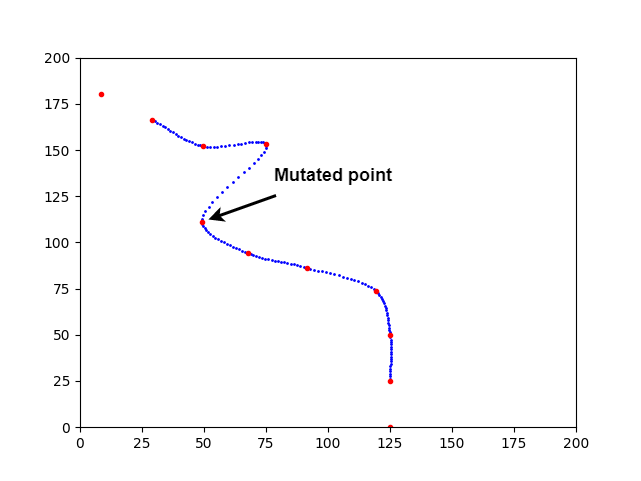}
         \caption{Mutated road }
         \label{fig:Mutated_Road}
     \end{subfigure}
        \caption{Mutation operation on the road}
        \label{fig:Mutation}
\end{figure}

\textit{Mutation:} The mutation operation also targets the coordinate values of the control points. We use \textit{Polynomial Bounded mutation}, a bounded mutation operation for real-valued individuals which was used in NSGA-II \cite{deb2002fast}. It features using a polynomial function for the probability distribution and a user-set parameter, $\eta$, presenting the \textit{crowding degree} of the mutation and adjusting the diversity in the resulting mutant. A high value for $\eta$ results in a mutant resembling the original solution, while a low $\eta$ leads to a more divergent mutant from the original.  
This domain-specific polynomial bounded mutation operation selects randomly a point---mutation point---in the set of control points and mutates randomly the $x$ or $y$ coordinate of the selected control point (see Figure \ref{fig:Mutation}). Accordingly, the sample points are re-calculated for the mutated set of control points, and their validity w.r.t the geometrical constraints are checked. In case the mutant does not represent a valid road, another control point is tried. 

\pink{It is also worth noting that the rates of the crossover and mutation have been set empirically and with regard to the characteristics of individuals in the problem. According to our empirical assessment, due to the existing inherent similarity between the individuals, the crossover operation could not lead to a big impact. Basically, since the test roads, as the individuals, are subject to geometrical and domain-specific constraints, the valid generated roads over the generations, do not show a high level of variance commonly, so the resulting roads from the crossover operation are very likely to be either similar to the parents or not valid at all. Then, in this regard, the mutation can take a more prominent role and have a better impact, and based on these observations, we set the crossover and mutation rates to 0.3 and 0.7 respectively.} The GA-driven test generator of {Deeper} is configured as presented in Algorithm \ref{Algorithm:GA}.

\begin{algorithm} [h]
\begin{flushleft}
\caption{GA-driven test generator in Deeper}\label{Algorithm:GA}
\begin{algorithmic}
1. Initialize a population of test scenarios (with $size =70$) from the quality population seed\\
\noindent 2. Evaluate the test scenarios through rendering them in the simulation and computing the fitness values\\
\Repeat {reaching the end of the test budget (e.g., given time)}{
3. Select highly-valued scenarios using tournament selection ($Tournament\ size =3$)\\
4. Create offspring by using crossover and mutation operations\\
     \quad 4.1. Apply the domain-specific crossover operation ($Crossover\ rate = 0.3$) \\
      \quad 4.2. Apply the domain-specific polynomial Bounded mutation operation $(Mutation\ rate =0.7)$ \\
5. Evaluate the offspring }
6. Collect the test scenarios revealing OBE failures\\
\end{algorithmic}
\end{flushleft}
\end{algorithm}

\subsubsection{\textbf{Evolution Strategies}} \label{D:Sec::Search_Algorithms_ES} \blue{The ES-driven test generators in Deeper use two canonical $(\mu + \lambda)$ and $(\mu, \lambda)$ ES algorithms.}
\blue{Algorithm \ref{Algorithm:mu+lambda} presents the procedure of $(\mu + \lambda)$ ES-driven test generator, where $\mu=70$ and $\lambda =30$. The procedure of $(\mu,\lambda)$ ES test generator is almost the same as $(\mu + \lambda)$ ES, and the main difference is in the selection step, where the $\mu$ highly-valued individuals are selected only from the offspring in $(\mu,\lambda)$ ES.
However, with regard to the required convergence condition $(\mu < \lambda)$ in this case, then those parameters in $(\mu,\lambda)$ ES test generator are configured as $\mu=70$ and $\lambda =100$ (the size of the main population is fixed at both ES-driven test generators).}

\blue{Moreover in order to draw a better distinction between GA and ES algorithms in detail, it is worth noting that in Deeper both GA and ES use the same size for the main (original) population. However, in Es, we have a temporary population with size $\lambda$, which forms the offspring--it is the pool where the mutation and crossover operations are applied, while in GA, the original population (with size $\mu$) is used to form the offspring. Additionally, the way the  mutation and crossover operations are applied to the offspring's individuals are different. For instance, in GA, each individual of the offspring will be considered to undergo both mutation and crossover operations, i.e., both mutation and crossover will be separately tried for each individual and their occurrences will be based on the user-defined probabilities. While, in our ES-based algorithms, either crossover or mutation is applied to each individual of the offspring, i.e., each offspring member is given one chance and it would undergo either crossover or mutation, though in ES-based algorithms, mutation takes a more prominent role.}

\begin{algorithm}
\begin{flushleft}
\caption{$(\mu + \lambda)$ ES-driven test generator in Deeper }\label{Algorithm:mu+lambda}
\begin{algorithmic}
{1. \blue{Initialize a population P with size $\mu$ of test scenarios sampled from the quality population seed ($\mu=70$)}}\\
{2. Evaluate the test scenarios in the population P through simulation and computing the fitness values}\\
\Repeat {reaching the end of the test budget (e.g., given time)}{
{3. Create a temporary population $P_{T}$ with size $\lambda$ by reproduction of test scenarios from the population P}\\
{4. Create an offspring by applying the crossover or mutation operation to the test scenarios in the population $P_{T}$ (with crossover probability $Cx_P = 0.3$ and mutation probability $Mu_P = 0.7$)}\\
{\quad 4.1. $choice = Random.random(),\ 0 < choice < 1$}\\
{ \quad 4.2. If $choice < Cx_P$ then}\\
{ \qquad Apply the domain-specific crossover operation}\\
{ \quad else if $choice < Cx_P + Mu_P$}\\
{\qquad Apply the domain-specific polynomial Bounded mutation operation}\\
{5. Evaluate the offspring}\\
{6. Select $\mu$ highly-valued test scenarios using tournament selection ($Tournament\ size =3$) from the original population P and the offspring}}
7. Collect the test scenarios revealing OBE failures
\end{algorithmic}
\end{flushleft}
\end{algorithm}

\subsubsection{\textbf{Particle Swarm Optimization}} \label{D:Sec::Search_Algorithms_PSO} The PSO-driven test generator in {Deeper} starts with initializing a population of particles from the quality population seed. Each test scenario is modeled as a particle, and the set of control points represents the position vector of the particle that is updated according to Eq. (\ref{eq:position_update}) and (\ref{eq:Velocity_update}) over various generations. After each updating, the corresponding sample points for the updated set of control points are calculated and the validity of the new set of road points is checked against the geometrical constraints. The PSO-driven test generator is configured as presented in Algorithm \ref{Algorithm:PSO} and based on the following setting for the parameters, which is $w=0.8$, $c1=2.0$, and $c2=2.0$.

\begin{algorithm}[h]
\begin{flushleft}
\caption{PSO-driven test generator in {Deeper}}\label{Algorithm:PSO}
\begin{algorithmic}
{1. Initialize a swarm of test scenario particles (with $size=70$) from the quality population seed}\\
{2. Evaluate the particles of the swarm through simulation and computing the fitness values}\\
{3. Select the global best particle w.r.t the fitness value $(G_{best})$}\\
\Repeat {reaching the end of the test budget (e.g., given time)} {
        \For {each test scenario particle P in the swarm}{ 
        {4. Calculate the particle's velocity according to eq. (2)}\\
        {5. Update the particle's position according to eq. (1)}\\
        {6. Evaluate the particle based on the fitness function}\\
        \If {fitness value of P is better than the local best of P, ($P_{best}$),}{
        {7. Update $P_{best}$ with $P$}
        }
        \If {fitness value of P is better than the global best, ($G_{best}$),}{
        {8. Update $G_{best}$ with $P$}
        }
        }
        }
9. Collect the test scenario particles revealing OBE failures
\end{algorithmic}
\end{flushleft}
\end{algorithm}



\section{Empirical Evaluation} \label{D:sec::EmpiricalEvaluation}
We conduct an empirical evaluation of the proposed simulation-integrated bio-inspired test generators in {Deeper}, by running experiments on an experimental setup based on a PC with 64-bit Windows 10 Pro, Intel Core i7-8550U CPU @ 1.80GHz, 16GB RAM, Intel UHD Graphics 620, and BeamNG.tech driving simulator together with the software requirements for running {Deeper}\footnote{see requirements at \url{https://github.com/mahshidhelali/Deeper_ADAS_Test_Generator.git}}.

\textit{Test Subject:} The system under test is BeamNG's
built-in driving agent, BeamNG.AI. It is an autonomous agent utilizing optimization techniques to plan the driving trajectory according to the speed limit while keeping the ego car inside the road lane. It is equipped with a DNN-based lane-keeping ADAS. The DNN-based lane-keeping system learns a mapping from the input of the onboard camera in the simulated environment to the steering angle. It is based on the \textsc{DAVE-2} architecture including a normalization layer, five convolutional layers followed by three fully connected layers \cite{bojarski2016end}. This test subject has been used in previous research and also in the SBST 2021 cyber-physical tool competition for evaluating test scenario generators \cite{gambi2019asfault, riccio2020model, panichella2021sbst}, and moreover does not require manual training, which can mitigate the threats to the validity of the results \cite{panichella2021sbst}.     

\subsection{Research Method} \label{D:sec::Research Method}
We design and implement a set of experiments to answer the research questions:
\begin{enumerate}
    \item RQ1: How capable are these test generators to trigger failures?
    \item RQ2: How diverse are the generated failure-revealing test scenarios?
    \item RQ3: How effectively and efficiently do the test generators perform? \blue{i.e., given a certain test budget, 3.1) how many test scenarios are generated, 3.2) what proportion of the scenarios is valid, and finally 3.3) what proportion of the valid test scenarios leads to triggering failures?}
\end{enumerate}

\begin{figure}
  \centering
  \includegraphics[width=.85\columnwidth]{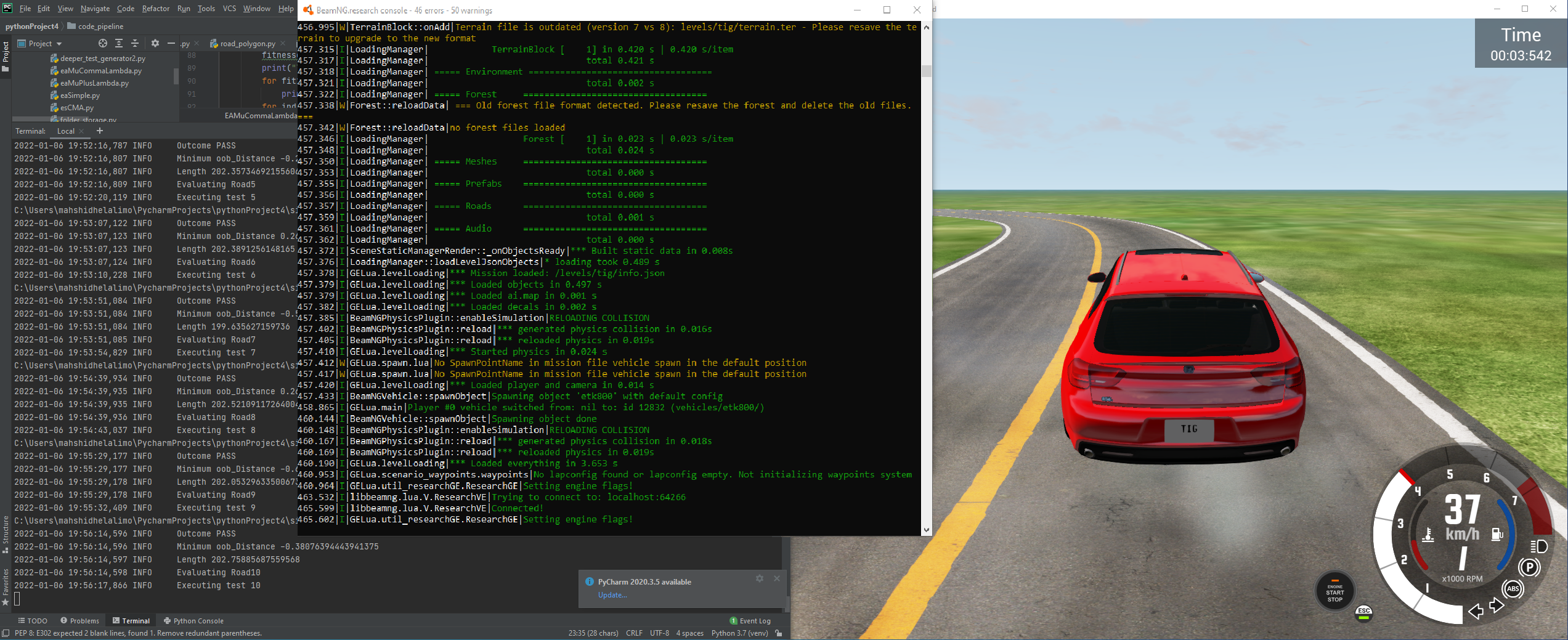}
  \caption{An overview of the experimental setup}
  \label{fig:Experimental_Setup}
\end{figure}

The experiments are simulation scenarios generated by a Python test scenario generator and executed by the simulation engine. BeamNG.AI is the autonomous driving agent controlling the ego car in the simulation (Figure \ref{fig:Experimental_Setup}). In order to provide quantitative answers to the RQs, we use the following quality criteria to assess the bio-inspired test scenario generators in {Deeper}:
\begin{itemize}
    \item Detected Failures: The number of generated test scenarios that lead to failures, w.r.t the given tolerance threshold.
    \item Failure Diversity: The dissimilarity between the test scenarios that lead to the failures. Generating diverse failure-revealing test scenarios is of interest, since triggering the same failures multiple times results in wasting the test budget, e.g., computation resources. In order to measure the failure diversity, we rely on a two-step strategy adopted by the SBST 2021 tool competition. It extracts, first, the road segments related to the failures, then computes the \textit{sparseness}, which is considered as the average of the maximum Levenshtein distance \cite{levenshtein1966binary} between those road segments.\\
    The failure-related road segments are referred to as the parts of the road 30 meters before the OBE and 30 meters after it, and accordingly, the sparseness is calculated as follows:
    \begin{equation}
     Sparseness = \frac{\sum_{i\in OBEs}\max_{j\in OBEs} Lev\_dist(i, j)}{|OBEs|}
    \end{equation}
    where $Lev\_dist(i, j)$ indicates the weighted Levenshtein distance between the road segments.
    \item Test generation efficiency and effectiveness: It indicates how the test generator uses the given test budget to generate the test scenarios, in particular, how many test scenarios are generated in total, what fraction of them are valid, and what fraction of the valid ones triggers failures.
\end{itemize}

\textit{Experiments:} We design two sets of experiments as implemented in the SBST 2021 CPS testing tool competition. In order to provide a comparative analysis, we compare the results of the proposed test generators in {Deeper} with the presented test generators in the tool competition, i.e., Frenetic \cite{Frenetic}, GABExploit and GABExplore \cite{GABezier}, Swat \cite{SWAT}, and also the earlier version of {Deeper} \cite{Moghadam2021Deeper}. We run the test generators on the test subject based on the same two experiment configurations as in the competition, which are shown in table \ref{table:Experiment_configuration}. The \textit{SET1} of experiments provides a 5-hour test generation budget, meanwhile sets the failing tolerance threshold to a high value, $0.95$, and does not consider any speed limit. This experiment configuration might lead to a more careless style of driving. The \textit{SET2} of the experiments allocates a shorter time budget for the test generation and considers a lower tolerance threshold, $0.85$, while imposing a speed limit of 70 km/h---promoting a more careful driving style. To ensure a fair comparison, we run each tool the same number of times; we run each test generator $5$ times in the experiment configuration SET1 and $10$ times in SET2 and report distributions of the results. \pink{It is to be noted that due to the long execution time of each simulation run and more importantly in order to make a fair comparison between the newly proposed test generators and the counterpart tools---w.r.t addressing the same problem, under the same evaluation conditions, and for answering the same research questions---the number of runs for each tool in the experiment configurations has been tied to the setup used in SBST Tool Competition 2021 and the results for the counterpart tools in the empirical evaluation also rely on the published results by the competition. }


\pink{Furthermore, in order to analyze and provide a better insight into the statistics of the results, we run the Kruskal-Wallis test \cite{kruskal1952use} on the results in each category, i.e., number of detected failures, sparseness, and effectiveness plus. Here, for each analysis, we have 9 groups of data, each of which represents a test generator. Kruskal-Wallis is a non-parametric test and an extension of the Mann-Whitney U test \cite{wilcoxon1992individual} (also known as Wilcoxon Rank-Sum test) for comparing more than two groups of data---to evaluate the hypothesis that the populations, from which the groups are sampled, are equivalent. This implies that none of the group populations is dominant over the others. A group is considered dominant if it has a higher probability of containing the largest element when one element is randomly selected from each of the group populations. The Kruskal-Wallis test is essentially used when the  sizes of the data groups are small and no assumption about the underlying data distribution is made. Kruskal-Wallis test reports the test statistic, H statistic, and the corresponding p-value.} 

\pink{Kruskal-Wallis test detects a significant difference between the data groups overall but does not state which group is different. Then, a pairwise comparison can be utilized to identify the specific differences between the data groups. Pairwise Mann-Whitney U test is one of the appropriate follow-up pairwise tests to be done after the Kruskal-Wallis test. It does not rely on assumptions about the population parameters, e.g., normality or equal variances, either. It is based on the ranks of the data rather than the actual values. In Mann-Whitney U test the null hypothesis is that there is no big difference between the two intended data groups and the distributions of the data groups are equivalent. The test is performed by ranking the observations from lowest to highest, combining the ranks from both groups and calculating the sum of the ranks per each group. The Mann-Whitney U test reports the U statistics corresponding with each data group and the p-value accordingly. The U statistics are the sums of the ranks of the observations in each group and the smaller U statistic is chosen as the Mann-Whitney test statistic and is used for calculating the p-value.
Here, the associated significance level ($\alpha$) is chosen as $ \alpha = 0.05$. A smaller p-value indicates stronger evidence against the null hypothesis, so If the p-value is less than or equal to the significance level, the null hypothesis is rejected, otherwise, the test fails to reject it.}

\begin{table}[h]
\centering
\caption{Experiment configurations}
\begin{tabular}{ |m{3.2cm}|m{2.8cm}|m{2.8cm}|m{2.8cm}|m{2.8cm}|}
\hline
\textbf{Name} & \textbf{Test Budget (h)} & \textbf{Map Size (m\textsuperscript{2})} & \textbf{Speed Limit (Km/h)} & \textbf{Failing Tolerance Threshold ($\%$)}\\
\hline
SET1 (careless driving) & 5h  & $200\times200$  & None & 0.95 \\
\hline
SET2 (cautious driving) & 2h  & $200\times200$  & 70 Km/h & 0.85 \\
\hline
\end{tabular}
\label{table:Experiment_configuration}
\end{table}


\section{Results and Discussion}\label{D:Sec::Result&Discussion}
This section reports results corresponding to the three RQs.

\subsection{Detected Failures (RQ1)}
\begin{figure}[h]
  \centering
   \begin{subfigure}[b]{0.49\textwidth}
         \centering
         \includegraphics[width=\textwidth, height =6cm]{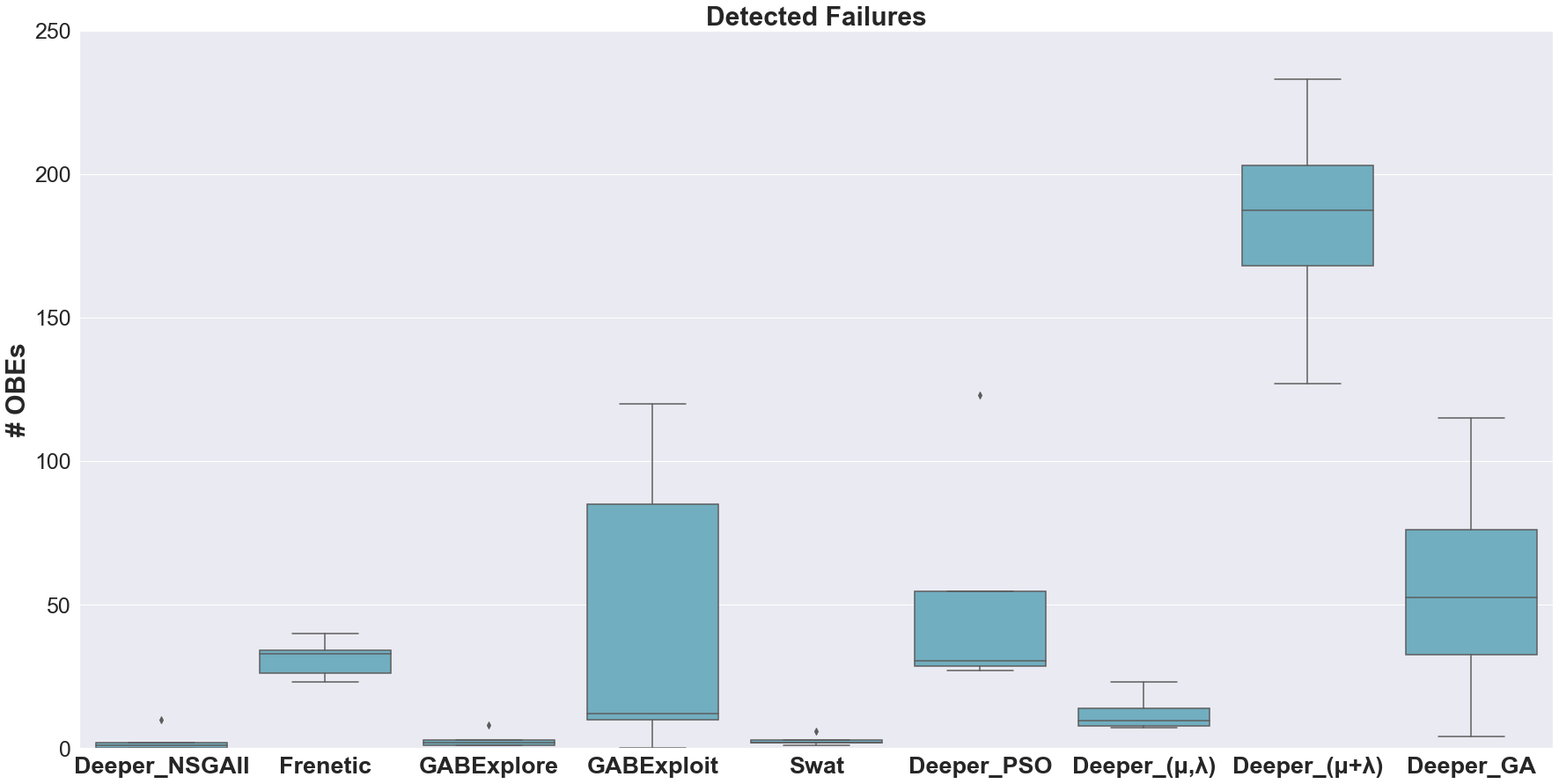}
         \caption{SET1}
         \label{fig:Exp_1_DetectedFailures}
     \end{subfigure}
     \begin{subfigure}[b]{0.49\textwidth}
         \centering
         \includegraphics[width=\textwidth,  height =6cm]{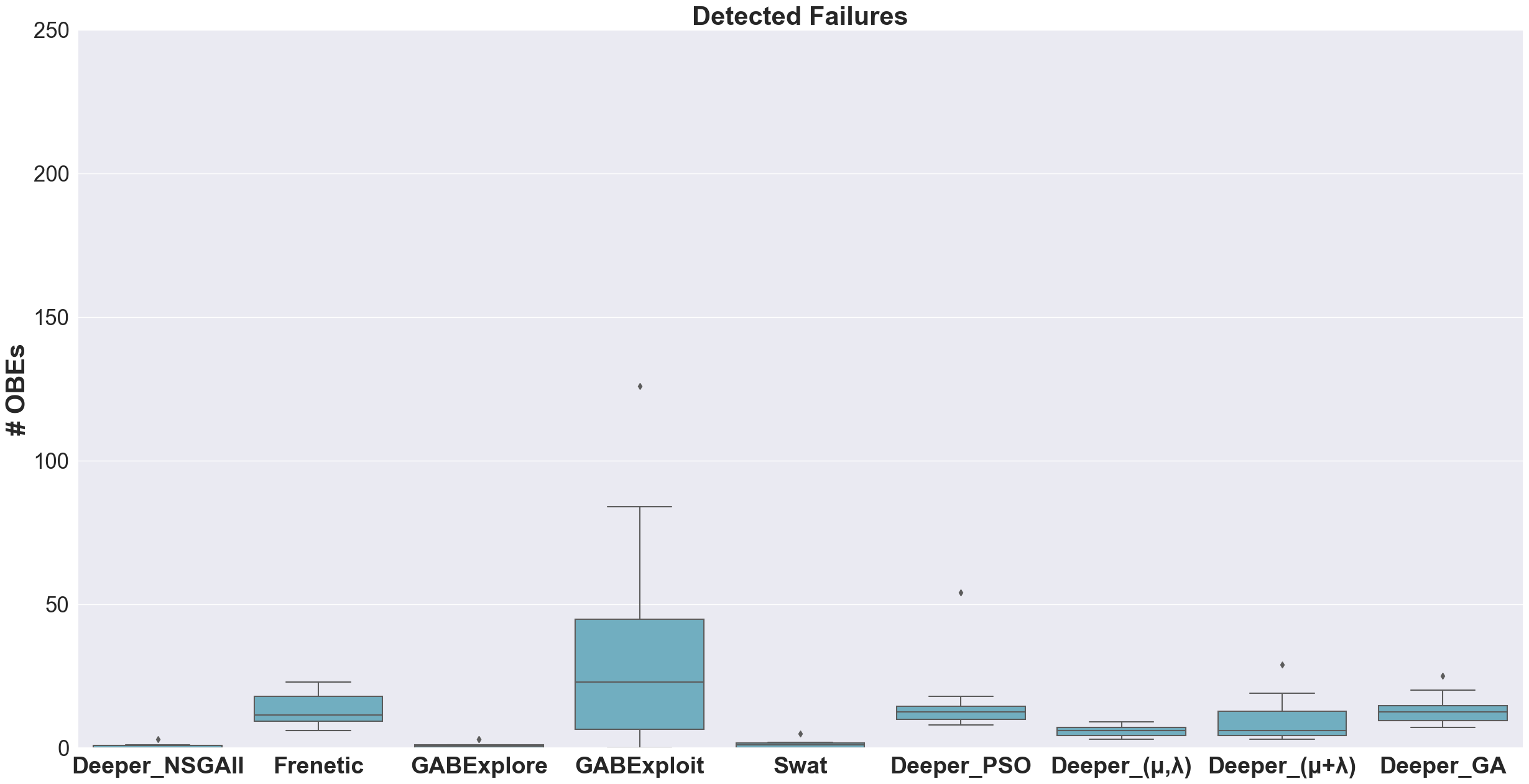}
         \caption{SET2}
         \label{fig:Exp_2_DetectedFailures}
     \end{subfigure}
\caption{\blue{Number of detected failures in SET1 and SET2}}
 \label{fig:DetectedFailures_SET1_SET2}
\end{figure}

\pink{In order to gain a statistical overview of the results related to the number of detected failures, first we run a Kruskal-Wallis statistical test on the groups of collected data from the test generators (see Table \ref{table:Kruskal-Wallis_SET1_SET2}). The p-values resulting from the Kruskal-Wallis test on SET1 and SET2 indicate significantly different behavior---in terms of the number of triggered failures---between the test generators in both experiment configurations. 
Moreover, in pursuance of achieving a deeper view of the difference between the test tools w.r.t number of detected failures, we perform the Mann-Whitney U test to show the pair-wise comparison between the test generators. According to the Mann-Whitney test results---shown in Table \ref{table:MannWhitney_DetectedFailures}---in a normal experimental condition such as SET1, {Deeper} $(\mu+\lambda)$ ES clearly behaves differently from all other tools 
(Table \ref{table:MannWhitney_DetectedFailures_SET1}). Meanwhile, the test does not prove a considerable difference in the comparison of Deeper PSO- and GA-driven test generators with either Frenetic or GABExploit---the top performers in the competition. In particular, the further boxplots (see Figure \ref{fig:Exp_1_DetectedFailures}) for Deeper PSO and GA actually show comparable performance---according to the median of the values---to Frenetic.
Regarding the behavior of the test tools in SET2---under strict conditions---we see that all the new test generators, Deeper PSO, $(\mu+\lambda)$, $(\mu, \lambda)$, and GA, show totally different behavior compared to Swat, GABExplore, and Deeper NSGA-II. Meanwhile, regarding Deeper PSO, $(\mu+\lambda)$, and GA, no considerably different behavior compared to either of Frenetic and GABExploit has been proved according to the Mann-Whitney test (see Table \ref{table:MannWhitney_DetectedFailures_SET2}). Similarly, the boxplots in Figure \ref{fig:Exp_2_DetectedFailures}, w.r.t the median of the number of triggered failures, for Deeper PSO and GA show quite analogous performance to Frenetic.}  

\blue{Along with the statistical tests, we also use boxplots to draw a picture of the distribution of the number of failures triggered by the test tools through indicating minimum, first quartile (Q1), median, third quartile (Q3) and maximum of the result data.} Figure \ref{fig:Exp_1_DetectedFailures} reports the number of triggered failures by each of the tools in experiment configuration SET1. In this regard, as shown in Figure \ref{fig:Exp_1_DetectedFailures}, $(\mu+\lambda)$ ES-driven test generator in {Deeper} could successfully trigger at least 2X more OBEs than the highest record---held by GABExploit---this significantly different behavior has been also indicated by the results of the statistical test. At the same time, {Deeper} $(\mu+\lambda)$ ES showed more consistent performance over the runs, i.e., with lower standard deviation, compared to GABExploit, which revealed highly varying behavior across the runs (e.g., returning over 100 OBEs in some runs, but failed to trigger any failure in other runs \cite{panichella2021sbst}). 
More importantly, the PSO- and GA-driven approaches were able to trigger the failures in which the ego car invades the opposite lane of the road---a type of failure that has been typically considered difficult to trigger in most of the test generator tools \cite{panichella2021sbst}. 


Similarly, Figure \ref{fig:Exp_2_DetectedFailures} presents the number of triggered failures in experiment configuration SET2. With regard to the limited time budget, the speed limit of 70 km/h and the tolerance threshold $0.85$, in contrast to {Deeper} NSGA-II, GABExplore, GABExploit, and Swat, none of the newly proposed test generators in {Deeper} left an experiment without triggering any failure or with a very low number of failures (i.e., less than $3$). It means that they are able to detect failures even within a limited test budget and strict constraints such as setting a speed limit (see Table \ref{table:TriggeredFailures_SET2}). It is noted that {Deeper} NSGA-II, GABExplore, and Swat triggered just equal or less than $1$ failure in a considerable number of experiments in SET2. Besides the wide range of number of detected failures by GABExploit, the PSO- and GA-driven test generators result in a comparable number of detected failures to Frenetic. Additionally, in the experiment configuration SET2, the GA-driven test generator was also able to trigger the failures showing the invasion of the car to the opposite lane of the road.

\begin{table}[h]
\centering
\caption{Triggered failures in SET2}
\begin{tabular}{ |m{2.8cm}|m{1cm}|m{1cm}|m{1cm}|m{1cm}|m{1cm}|m{1cm}|m{1cm}|m{1cm}|m{1cm}|m{1cm}|}
\hline
\textbf{Test Generator} & \textbf{Exp.1} & \textbf{Exp.2} & \textbf{Exp.3} & \textbf{Exp.4}&\textbf{Exp.5} &\textbf{Exp.6} &\textbf{Exp.7} &\textbf{Exp.8} &\textbf{Exp.9} &\textbf{Exp.10} \\
\hline
Deeper\_NSGA-II & \CL 0 & \CL 0 & \CL 0 & \CLL 1 & \CLL 1 & \CL 0 & \CL 0 & \CL 0 & 3 & \CL 0 \\
\hline
Frenetic & 12 & 8 & 9 & 11 & 19 & 10 & 15 & 20 & 6 & 23 \\
\hline
GABExplore & \CL 0 & \CLL 1 & \CLL 1 & \CL 0 & \CLL 1 & 3 & \CL 0 & \CL 0 & 3 & \CL 0 \\
\hline
GABExploit & 84 & 18 & 126 & 38 & \CL  0 & 11 & 47 & \CL 0 & 28 & 5 \\
\hline
Swat & \CL 0 & \CL 0 & \CL 0 & \CL 0 & 5 & \CLL 1 & \CLL 1 & 2 & \CLL 1 & 2 \\
\hline
Deeper\_PSO & 54 & 8 & 13 & 13 & 10 & 9 & 10 & 15 & 18 & 12 \\
\hline
Deeper\_$(\mu,\lambda)$ ES & 8 & 4 & 3 & 5 & 7 & 7 & 3 & 5 & 7 & 9 \\
\hline
Deeper\_$(\mu+\lambda)$ ES & 29 & 5 & 9 & 4 & 4 & 7 & 19 & 3 & 14 & 5 \\
\hline
Deeper\_GA & 11 & 7 & 11 & 15 & 14 & 9 & 25 & 9 & 14 & 20 \\
\hline
\end{tabular}
\label{table:TriggeredFailures_SET2}
\end{table}

\begin{table}[h]
\centering
\caption{\pink{Summary of Kruskal-Wallis test on the number of detected failures, sparseness, and effectiveness plus promoted by each test generator in SET1 and SET2}}
\begin{tabular}{|m{2.5cm}|m{1.5cm}|m{1.5cm}|m{1.5cm}|m{1.5cm}|}
\hline
\textbf{} & \multicolumn{2}{|c|}{\textbf{H statistic}} & \multicolumn{2}{|c|}{\textbf{p-value}} \\
\hline
\textbf{} & {SET1} & {SET2} & {SET1} & {SET2} \\
\hline
Detected Failures &  {33.045} & {60.090} & {6.0437e-05} & {4.4740e-10} \\
\hline
Sparseness & 21.300 & 25.734 & 0.0016  & 0.0001 \\
\hline
Effectiveness Plus & 33.180 & 60.115 & 5.7160e-05   & 4.4254e-10 \\
\hline
\end{tabular}
\label{table:Kruskal-Wallis_SET1_SET2}
\end{table}

\begin{table}[h]
\caption{\pink{Mann-Whitney U test on the detected failures in SET1 and SET2}}
    \begin{subtable}[h]{0.45\textwidth}
        \centering
        \caption{SET1}
        \begin{tabular}{|p{2cm}|p{2cm}|p{0.5cm}|p{0.5cm}|p{1cm}|}
        \toprule
        \textbf{Group1}       & \textbf{Group2}       & \textbf{U1} & \textbf{U2} & \textbf{p-value} \\ \midrule
        Deeper\_PSO           & Frenetic              & 13.0        & 12.0         & 1.0           \\
        \CLLL Deeper\_PSO           & GABExplore            &25.0   &0.0   &0.0079          \\
        Deeper\_PSO           & GABExploit            &17.0   &8.0    &0.4206           \\
        \CLLL Deeper\_PSO           & Swat                  &25.0   &0.0   &0.0079          \\
        \CLLL Deeper\_PSO           & Deeper\_NSGAII        &25.0   &0.0   &0.0079          \\
        \CLLL Deeper\_PSO           & Deeper\_$(\mu+\lambda)$  &0.0   &25.0   &0.0079         \\
        \CLLL Deeper\_PSO           & Deeper\_$(\mu,\lambda)$   &25.0   &0.0   &0.0079        \\
        Deeper\_PSO           & Deeper\_GA            &9.0   &16.0   &0.5476           \\
        \CLLL Deeper\_$(\mu+\lambda)$  & Frenetic              &25.0  &0.0   &0.0079           \\
        \CLLL Deeper\_$(\mu+\lambda)$  & GABExplore            &25.0   &0.0   &0.0079        \\
        \CLLL Deeper\_$(\mu+\lambda)$  & GABExploit            &25.0   &0.0   &0.0079         \\
        \CLLL Deeper\_$(\mu+\lambda)$  & Swat                  &25.0   &0.0   &0.0079         \\
        \CLLL Deeper\_$(\mu+\lambda)$  & Deeper\_NSGAII        &25.0   &0.0   &0.0079         \\
        \CLLL Deeper\_$(\mu+\lambda)$  & Deeper\_$(\mu,\lambda)$ &25.0   &0.0   &0.0079      \\
        \CLLL Deeper\_$(\mu+\lambda)$  & Deeper\_GA           &25.0   &0.0   &0.0079      \\
        \CLLL Deeper\_$(\mu,\lambda)$ & Frenetic              &0.5  &24.5   &0.0159    \\
        \CLLL Deeper\_$(\mu,\lambda)$ & GABExplore            &23.5   &1.5  &0.0317           \\
        Deeper\_$(\mu,\lambda)$ & GABExploit            &8.5  &16.5   &0.5476    \\
        \CLLL Deeper\_$(\mu,\lambda)$ & Swat                 &25.0   &0.0   &0.0079    \\
        Deeper\_$(\mu,\lambda)$ & Deeper\_NSGAII        &22.5  &2.5  &0.0556    \\
        Deeper\_$(\mu,\lambda)$ & Deeper\_GA            &5.0  &20.0   &0.1508   \\
        Deeper\_GA            & Frenetic              &20.0  &5.0   &0.1508   \\
        \CLLL Deeper\_GA            & GABExplore            &24.0   &1.0   &0.0159  \\
        Deeper\_GA            & GABExploit          &14.0   &11.0   &0.8413   \\
        \CLLL Deeper\_GA            & Swat                  &24.0   &1.0  &0.0159   \\
        \CLLL Deeper\_GA            & Deeper\_NSGAII        &24.0  &1.0  &0.0159 \\
        Deeper\_NSGAII        & Frenetic              &0.0  &25.0   &0.0079  \\
        Deeper\_NSGAII        & GABExplore            &8.5  &16.5   &0.5476 \\
        Deeper\_NSGAII        & GABExploit            &4.5  &20.5   &0.1508  \\
        Deeper\_NSGAII        & Swat                  &7.5  &17.5   &0.4206  \\
        Frenetic              & GABExplore            &25.0   &0.0   &0.0079 \\
        Frenetic              & GABExploit            &15.0  &10.0  &0.6905   \\
        Frenetic              & Swat                  &25.0   &0.0   &0.0079  \\
        GABExplore            & GABExploit            &5.0  &20.0   &0.1508 \\
        GABExplore            & Swat                  &11.5   &13.5   &1.0  \\
        GABExploit            & Swat                  &20.0   &5.0  &0.1508 \\ \bottomrule
        \end{tabular}
       \label{table:MannWhitney_DetectedFailures_SET1}
    \end{subtable}
    \hspace{1cm}
    \begin{subtable}[h]{0.45\textwidth}
        \centering
        \caption{SET2}
        \begin{tabular}{|p{2cm}|p{2cm}|p{0.5cm}|p{0.5cm}|p{1cm}|}
        \toprule
        \textbf{Group1}       & \textbf{Group2}       & \textbf{U1} & \textbf{U2} & \textbf{p-value} \\ \midrule
        Deeper\_PSO           & Frenetic              & 52.0        & 48.0        & 0.9118           \\
        \CLLL Deeper\_PSO           & GABExplore            & 100.0       & 0.0         & 0.0              \\
        Deeper\_PSO           & GABExploit            & 40.5        & 59.5        & 0.5288           \\
        \CLLL Deeper\_PSO           & Swat                  & 100.0       & 0.0         & 0.0              \\
        \CLLL Deeper\_PSO           & Deeper\_NSGAII        & 100.0       & 0.0         & 0.0              \\
        Deeper\_PSO           & Deeper\_$(\mu+\lambda)$  & 73.5        & 26.5        & 0.0892           \\
        \CLLL Deeper\_PSO           & Deeper\_$(\mu,\lambda)$ & 98.0        & 2.0         & 0.0              \\
        Deeper\_PSO           & Deeper\_GA            & 49.5        & 50.5        & 1.0              \\
        Deeper\_$(\mu+\lambda)$  & Frenetic              & 27.0        & 73.0    & 0.0892           \\
        \CLLL Deeper\_$(\mu+\lambda)$  & GABExplore            & 99.0        & 1.0         & 0.0              \\
        Deeper\_$(\mu+\lambda)$  & GABExploit            & 32.0        & 68.0        & 0.1903           \\
        \CLLL Deeper\_$(\mu+\lambda)$  & Swat                  & 96.0        & 4.0         & 0.0001           \\
        \CLLL Deeper\_$(\mu+\lambda)$  & Deeper\_NSGAII        & 99.5        & 0.5         & 0.0              \\
        Deeper\_$(\mu+\lambda)$  & Deeper\_$(\mu,\lambda)$ & 60.0        & 40.0        & 0.4813           \\
        Deeper\_$(\mu+\lambda)$  & Deeper\_GA            & 26.5        & 73.5        & 0.0892           \\
        \CLLL Deeper\_$(\mu,\lambda)$ & Frenetic              & 7.0         & 93.0        & 0.0005           \\
        \CLLL Deeper\_$(\mu,\lambda)$ & GABExplore            & 98.0        & 2.0         & 0.0              \\
        Deeper\_$(\mu,\lambda)$ & GABExploit            & 26.0        & 74.0        & 0.0753           \\
        \CLLL Deeper\_$(\mu,\lambda)$ & Swat                  & 96.0        & 4.0         & 0.0001           \\
        \CLLL Deeper\_$(\mu,\lambda)$ & Deeper\_NSGAII        & 99.0        & 1.0         & 0.0              \\
        \CLLL Deeper\_$(\mu,\lambda)$ & Deeper\_GA            & 4.5         & 95.5        & 0.0002           \\
        Deeper\_GA            & Frenetic              & 52.0        & 48.0        & 0.9118           \\
        \CLLL Deeper\_GA            & GABExplore            & 100.0       & 0.0         & 0.0              \\
        Deeper\_GA            & GABExploit            & 38.0        & 62.0        & 0.393            \\
        \CLLL Deeper\_GA            & Swat                  & 100.0       & 0.0         & 0.0              \\
        \CLLL Deeper\_GA            & Deeper\_NSGAII        & 100.0       & 0.0         & 0.0              \\
        Deeper\_NSGAII        & Frenetic              & 0.0         & 100.0       & 0.0              \\
        Deeper\_NSGAII        & GABExplore            & 39.5        & 60.5        & 0.4813           \\
        Deeper\_NSGAII        & GABExploit            & 13.0        & 87.0        & 0.0039           \\
        Deeper\_NSGAII        & Swat                  & 34.0        & 66.0        & 0.2475           \\
        Frenetic              & GABExplore            & 100.0       & 0.0         & 0.0              \\
        Frenetic              & GABExploit            & 38.5        & 61.5        & 0.4359           \\
        Frenetic              & Swat                  & 100.0       & 0.0         & 0.0              \\
        GABExplore            & GABExploit            & 15.0        & 85.0        & 0.0068           \\
        GABExplore            & Swat                  & 44.5        & 55.5        & 0.7394           \\
        GABExploit            & Swat                  & 83.5        & 16.5        & 0.0115         \\ \bottomrule 
        \end{tabular}
        \label{table:MannWhitney_DetectedFailures_SET2}
     \end{subtable}
     \label{table:MannWhitney_DetectedFailures}
\end{table}

\subsection{Diversity of Failures (RQ2)}
\begin{figure}[h]
  \centering
   \begin{subfigure}[b]{0.49\textwidth}
         \centering
         \includegraphics[width=\textwidth, height =6cm]{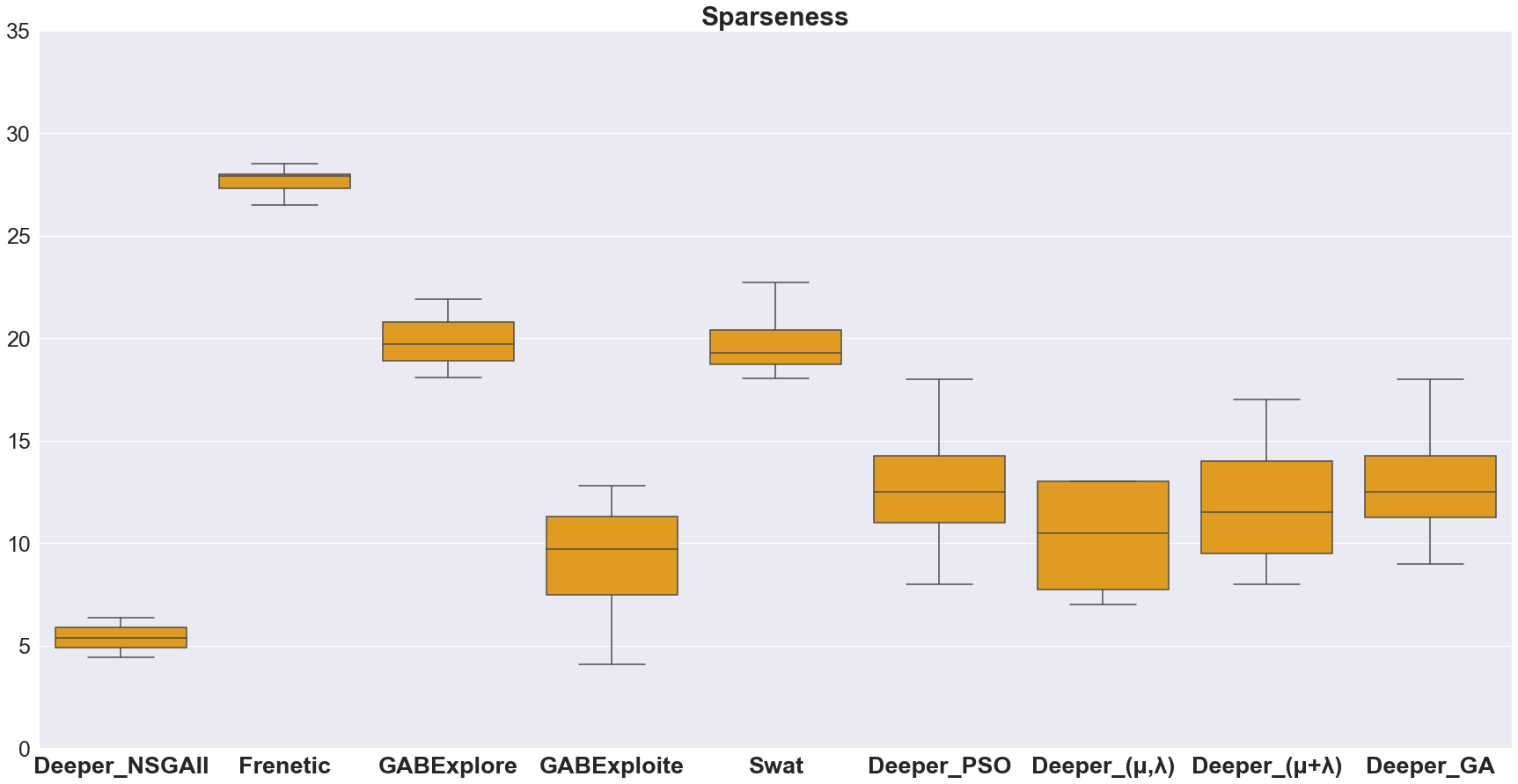}
         \caption{SET1}
         \label{fig:Exp_1_FailureDiversity}
     \end{subfigure}
     \begin{subfigure}[b]{0.49\textwidth}
         \centering
         \includegraphics[width=\textwidth,  height =6cm]{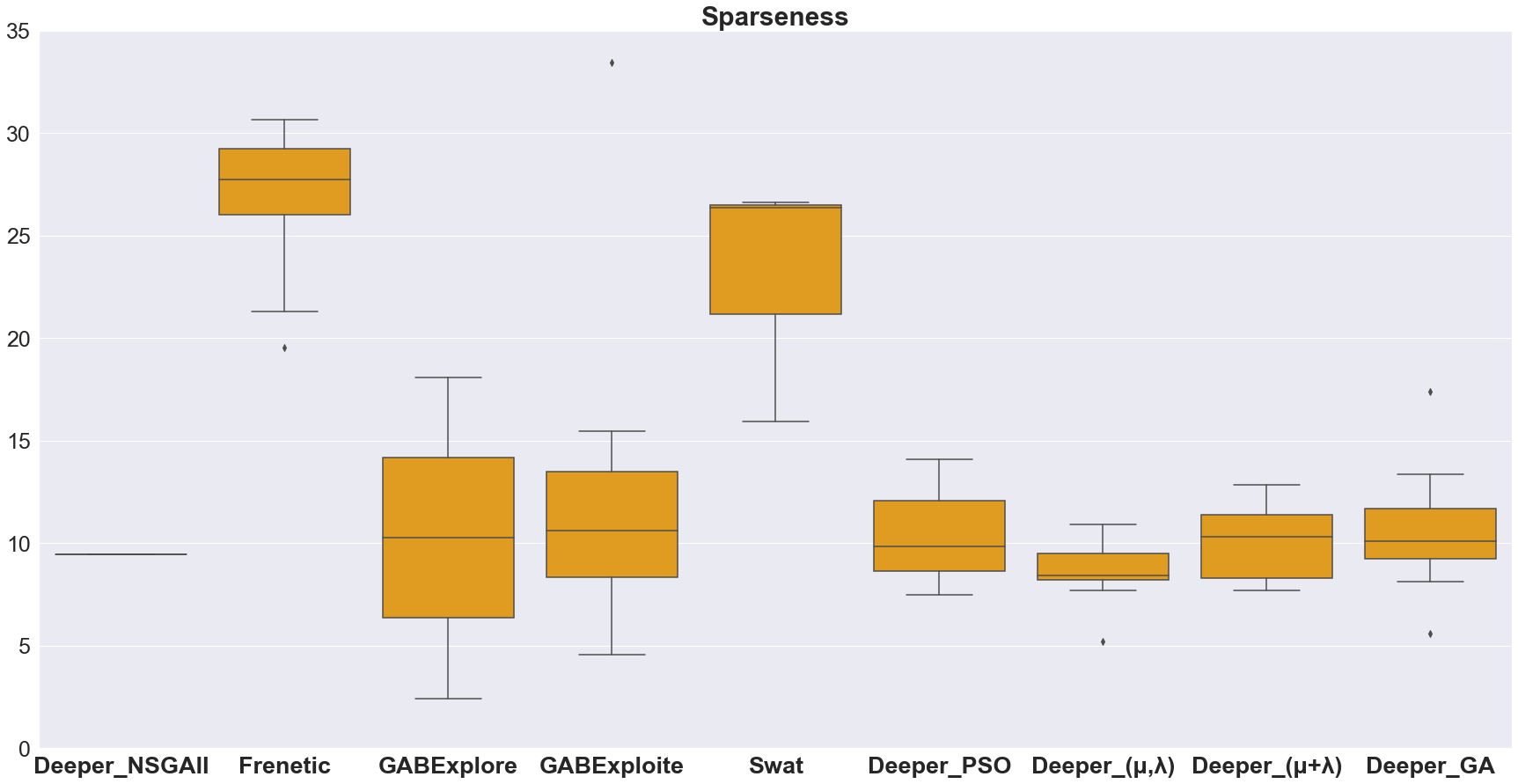}
         \caption{SET2}
         \label{fig:Exp_2_FailureDiversity}
     \end{subfigure}
\caption{\blue{Failure diversity in terms of sparseness in SET1 and SET2}}
 \label{fig:FailureDiversity_SET1_SET2}
\end{figure}

\pink{Tables \ref{table:Kruskal-Wallis_SET1_SET2} 
and \ref{table:MannWhitney_sparseness} present the results of the Kruskal-Wallis and pair-wise Mann-Whitney U statistical tests on the data showing the promoted failure diversity by the test generators respectively. It is important to note that in the statistical tests on the sparseness data, the tools that did not report sparseness value in more than $20\%$ of the runs---due to the very low number of triggered failures, i.e., zero or one---were excluded from the tests, since the basic assumptions for the possibility of doing the test, would be distorted. In particular, Deeper NSGA-II and GABExplore were excluded in the statistical analysis of sparseness data in SET1, and in addition to these tools, Swat was also kept out in the statistical tests on the sparseness data from SET2. As shown by the Mann-Whitney U test on the sparseness data from SET1 (see Table \ref{table:MannWhitney_sparseness_SET1}), Frenetic and Swat perform differently from the rest of the tools, and notable differences between the behavior of the Deeper test generators and either Frenetic or Swat have been reported by the test. However, regarding the comparison between the new test generators in {Deeper} and GABExploit, the test does not prove a significant difference in their behavior. Later, the boxplots (see Figure \ref{fig:Exp_1_FailureDiversity}) actually, for new Deeper test generators, show comparable performance, w.r.t the median of the sparseness values, to GABExploit.} 

\pink{In strict experimental conditions (SET2), the Mann-Whitney test (see Table \ref{table:MannWhitney_sparseness_SET2}) indicates that Frenetic show different behavior in terms of promoted diversity from the Deeper test generators. It is also worth noting that regarding SET2 all the new Deeper test generators are included in the statistical analysis, since they all triggered enough failures in all runs. Though, among the rest of the tools, only Frenetic and GABExploit were eligible, and the other ones were excluded due to the fact that in a considerable number of runs under SET2, they did not trigger enough failures to calculate the sparseness.}

\begin{table}[h]
\caption{\pink{Mann-Whitney U test on the sparseness (failure diversity) in SET1 and SET2}}
    \begin{subtable}[h]{0.45\textwidth}
        \centering
        \caption{SET1}
        \begin{tabular}{|p{2cm}|p{2cm}|p{0.5cm}|p{0.5cm}|p{1cm}|}
        \toprule
        \textbf{Group1}       & \textbf{Group2}       & \textbf{U1} & \textbf{U2} & \textbf{p-value} \\ \midrule
        \CLLL Deeper\_PSO           & Frenetic              &0.0 &25.0  &0.0079  \\
        Deeper\_PSO           & GABExploit            & 15.0 &5.0  &0.2857  \\
        \CLLL Deeper\_PSO           & Swat                  & 0.0 &20.0  &0.0159 \\
        Deeper\_PSO           & Deeper\_$(\mu+\lambda)$  & 14.0 &11.0  &0.8413  \\
        Deeper\_PSO           & Deeper\_$(\mu,\lambda)$ & 16.5 &8.5  &0.5476  \\
        Deeper\_PSO           & Deeper\_GA            & 12.0 &13.0  &1.0     \\
        \CLLL Deeper\_$(\mu+\lambda)$  & Frenetic              & 0.0 &25.0   &0.0079 \\
        Deeper\_$(\mu+\lambda)$  & GABExploit            & 14.0 &6.0  &0.4127  \\
        \CLLL Deeper\_$(\mu+\lambda)$  & Swat                  & 0.0 &20.0  &0.0159  \\
        Deeper\_$(\mu+\lambda)$  & Deeper\_$(\mu,\lambda)$ & 16.0  &9.0   &0.5476  \\
        Deeper\_$(\mu+\lambda)$  & Deeper\_GA            & 10.5 &14.5  &0.8413  \\
        \CLLL Deeper\_$(\mu,\lambda)$ & Frenetic              & 0.0  &25.0  &0.0079 \\
        Deeper\_$(\mu,\lambda)$ & GABExploit            & 12.0 &8.0  &0.7302 \\
        \CLLL Deeper\_$(\mu,\lambda)$ & Swat                  & 0.0  &20.0  &0.0159 \\
        Deeper\_$(\mu,\lambda)$ & Deeper\_GA            & 8.0 &17.0  &0.4206 \\
        \CLLL Deeper\_GA            & Frenetic              & 0.0 &25.0   &0.0079 \\
        Deeper\_GA            & GABExploit            & 16.0  &4.0  &0.1905   \\
        \CLLL Deeper\_GA            & Swat                  & 0.0  &20.0  &0.0159 \\
        Frenetic              & GABExploit            & 20.0 &0.0  &0.0159 \\
        Frenetic              & Swat                  & 20.0 &0.0  &0.0159  \\
        GABExploit            & Swat                  & 0.0 &16.0  &0.0286 \\ \bottomrule
        \end{tabular}
       \label{table:MannWhitney_sparseness_SET1}
    \end{subtable}
    \hspace{1cm}
    \begin{subtable}[h]{0.45\textwidth}
        \centering
        \caption{SET2}
        \begin{tabular}{|p{2cm}|p{2cm}|p{0.5cm}|p{0.5cm}|p{1cm}|}
        \toprule
        \textbf{Group1}       & \textbf{Group2}       & \textbf{U1} & \textbf{U2} & \textbf{p-value} \\ \midrule
        \CLLL Deeper\_PSO           & Frenetic              & 0.0         & 100.0       & 0.0              \\
        Deeper\_PSO                 & GABExploit            & 34.0          & 46.0      & 0.6334\\
        Deeper\_PSO           & Deeper\_$(\mu+\lambda)$  & 53.0        & 47.0        & 0.8534           \\
        Deeper\_PSO           & Deeper\_$(\mu,\lambda)$ & 72.0        & 28.0        & 0.1051           \\
        Deeper\_PSO           & Deeper\_GA            & 46.0        & 54.0        & 0.7959           \\
        \CLLL Deeper\_$(\mu+\lambda)$  & Frenetic              & 0.0         & 100.0       & 0.0              \\
        Deeper\_$(\mu+\lambda)$  & GABExploit   & 34.0    &46.0   &0.6334\\
        Deeper\_$(\mu+\lambda)$  & Deeper\_$(\mu,\lambda)$ & 68.5        & 31.5        & 0.1903           \\
        Deeper\_$(\mu+\lambda)$  & Deeper\_GA            & 44.0        & 56.0        & 0.6842           \\
        \CLLL Deeper\_$(\mu,\lambda)$ & Frenetic              & 0.0         & 100.0       & 0.0              \\
        Deeper\_$(\mu,\lambda)$  &GABExploit   & 25.0    &55.0    &0.2031\\
        Deeper\_$(\mu,\lambda)$ & Deeper\_GA            & 26.0        & 74.0        & 0.0753           \\
        \CLLL Deeper\_GA            & Frenetic              & 0.0        & 100.0        & 0.0   
              \\
        Deeper\_GA   &GABExploit   &38.0    &42.0    &0.8968 \\
        Frenetic     &GABExploit    &70.0   &10.0  &0.0062\\
         \bottomrule
        \end{tabular}
        \label{table:MannWhitney_sparseness_SET2}
     \end{subtable}
     \label{table:MannWhitney_sparseness}
\end{table}

Figure \ref{fig:Exp_1_FailureDiversity} depicts the diversity of the detected failures in SET1 in terms of the distribution of the failures' sparseness for all the test tools---including the ones that reported the sparseness in a  limited number of runs, i.e., GABExplore and Deeper NSGA-II. It is noted that GABExplore and Deeper NSGA-II in $40\%$ of the runs under SET1 did not report any sparseness figure, since they just triggered one or zero failure in each run. According to Figure \ref{fig:Exp_1_FailureDiversity}, all the new test generators in {Deeper} result in a considerable improvement in failure sparseness compared to the first version, {Deeper} NSGA-II, under normal experimental conditions (SET1). 
The newly proposed test generators also show comparable performance to GABExploit. 
Figure \ref{fig:Exp_2_FailureDiversity} shows the distribution of the failures' sparseness in SET2. In the limited test budget and strict driving constraints, all the newly test generators again show a big improvement in promoting failure sparseness in comparison to the first version, {Deeper} NSGA-II, under SET2. In the meantime, {Deeper} PSO, $(\mu+\lambda)$, and GA promote comparable levels of failure diversity to GABExploit---though in a more consistent way. At the same time, GABExplore and Swat in around $70\%$ of the experiments in SET2 did not provide any sparseness figure, since they triggered just one or zero failure (see Table \ref{table:TriggeredFailures_SET2}).

\subsection{Test Effectiveness and Efficiency (RQ3)}
\begin{figure}[h]
  \centering
   \begin{subfigure}[b]{0.49\textwidth}
         \centering
         \includegraphics[width=\textwidth, height =6cm]{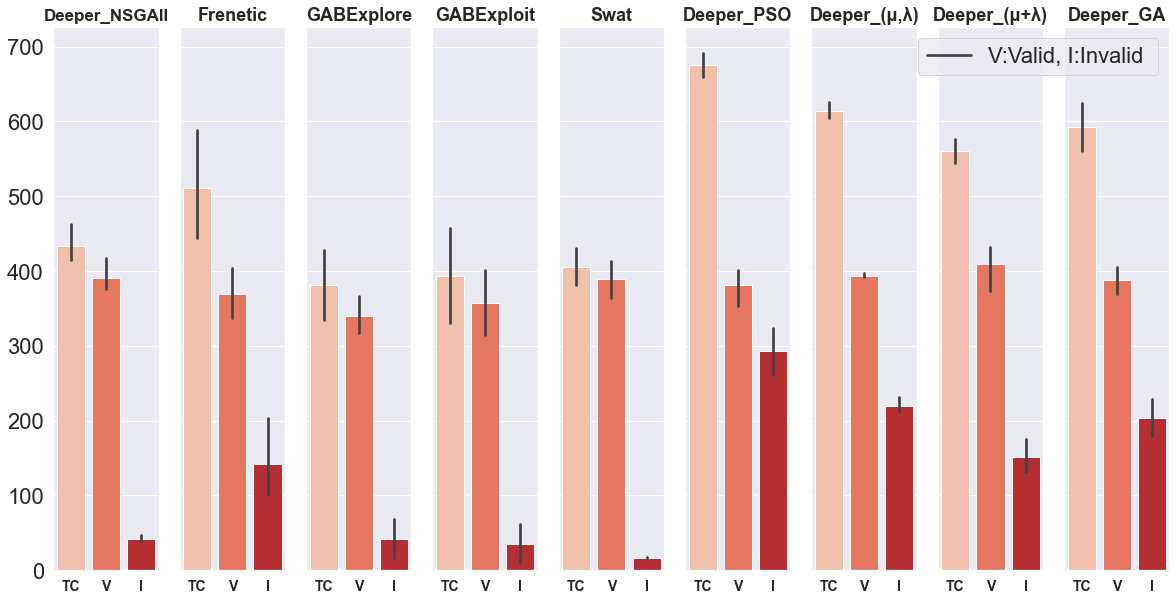}
         \caption{SET1}
         \label{fig:Exp_1_TestEfficiency}
     \end{subfigure}
     \begin{subfigure}[b]{0.49\textwidth}
         \centering
         \includegraphics[width=\textwidth,  height =6cm]{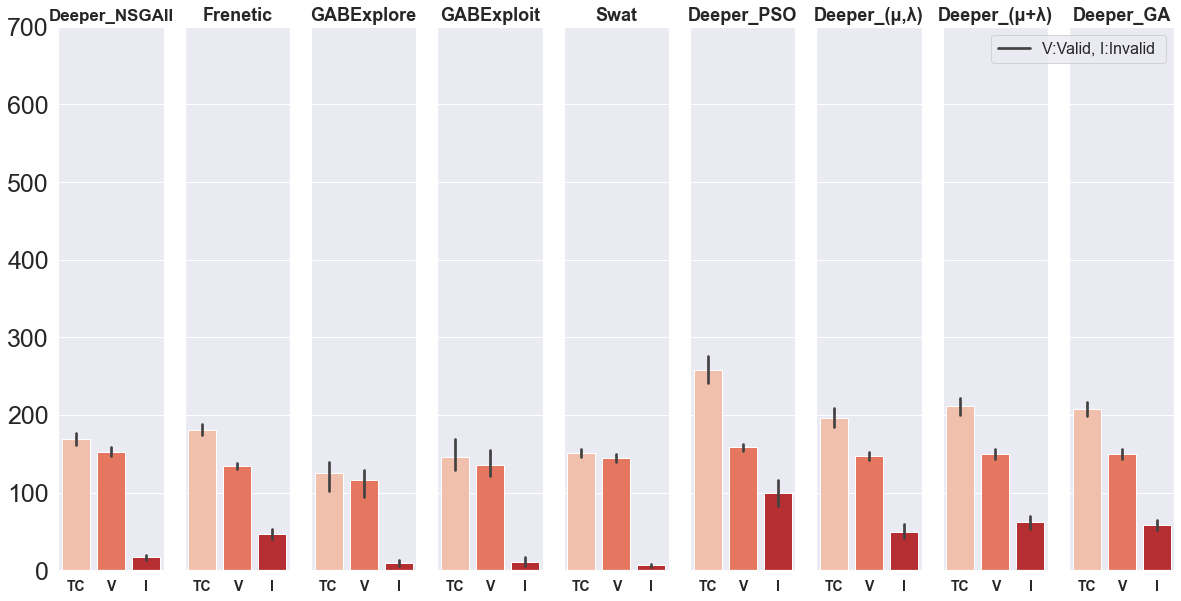}
         \caption{SET2}
         \label{fig:Exp_2_TestEfficiency}
     \end{subfigure}
\caption{\blue{Test generation effectiveness and efficiency in SET1 and SET2}}
 \label{fig:TestEfficiency_SET1_SET2}
\end{figure}

In SBST competition, test effectiveness and efficiency were indicated by how many test scenarios are generated and what proportion of the scenarios is valid, given a certain test budget. They basically show how well the test generator is able to utilize the test budget. Figures \ref{fig:Exp_1_TestEfficiency} and \ref{fig:Exp_2_TestEfficiency} report the average number of total test scenarios, as well as the number of valid and invalid scenarios generated by each tool in SET1 and SET2 respectively. Generally, the new test generators in {Deeper} utilize the test budget more efficiently than the competition tools and generate a higher number of test scenarios within the given test time. In this regard, {Deeper} PSO results in the highest efficiency (e.g., generates more than 650 scenarios on average within 5 hours) among all the test generators in both experimental configurations. Regarding the number of valid test scenarios, all the {Deeper} test generators along with Swat lead to an almost comparable number of valid scenarios. However, with respect to the ratio of the valid test scenarios to the total generated ones---called test effectiveness according to the competition evaluation---Swat, {Deeper} NSGA-II, and GABExploit are the ones showing the highest result.

To answer RQ3, in addition to the metrics defined and used by the competition, we define an additional metric, an aggregated test effectiveness metric called \textit{effectiveness plus}, which indicates what proportion of the valid test scenarios leads to triggering failures. It is defined as the ratio of the triggered failures to the number of valid test scenarios and is intended to present the effectiveness of the test generators w.r.t meeting the target---detecting failures. \pink{We conduct Kruskal-Wallis and pair-wise Mann-Whitney U statistical tests on the result data showing effectiveness plus as well (see Tables \ref{table:Kruskal-Wallis_SET1_SET2} and \ref{table:MannWhitney_effectiveness}). According to the results of the Mann-Whitney U test on the data, in non-strict experimental conditions (SET1), {Deeper} $(\mu+\lambda)$ ES-driven test generator shows a significantly different performance than other tools, meanwhile {Deeper} PSO and GA also prove a majorly different behavior than GABExplore, Swat, and Deeper NSGA-II (see Table \ref{table:MannWhitney_effectiveness_SET1}). Additionally, the Mann-Whitney U test does not prove a significant difference for the comparison of Deeper PSO or GA with either Frenetic or GABExploit. In SET2, the statistical test shows that GABExploit performs differently from others (see Table \ref{table:MannWhitney_effectiveness_SET2}), while again regarding the comparison of {Deeper} PSO, GA, and {Deeper} $(\mu+\lambda)$ ES with Frenetic no big difference has been proved by the Mann-Whitney U test .}

\begin{figure}[h]
  \centering
   \begin{subfigure}[b]{0.49\textwidth}
         \centering
         \includegraphics[width=\textwidth, height =6cm]{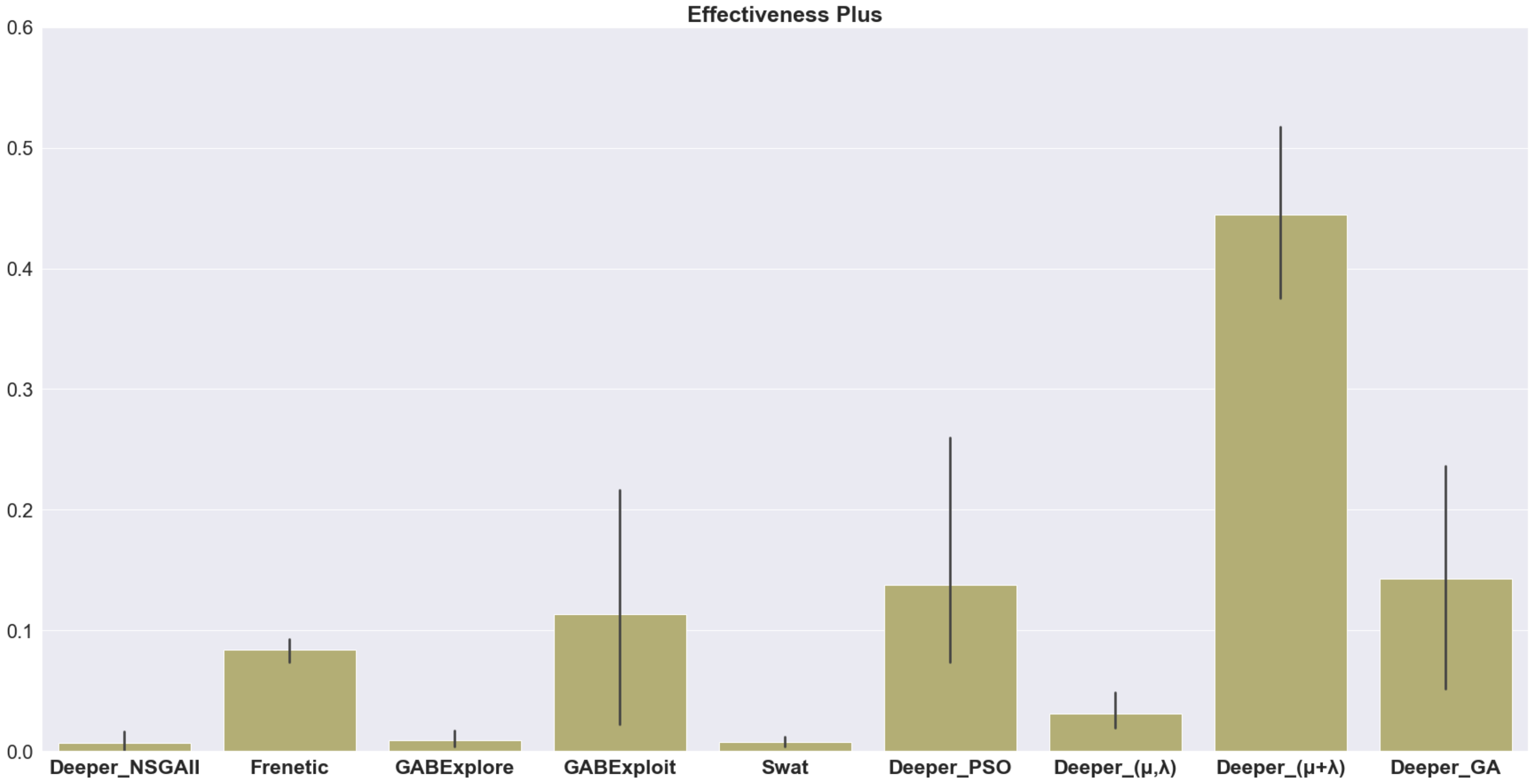}
         \caption{SET1}
         \label{fig:Exp_1_Effectiveness_Plus}
     \end{subfigure}
     \begin{subfigure}[b]{0.49\textwidth}
         \centering
         \includegraphics[width=\textwidth,  height =6cm]{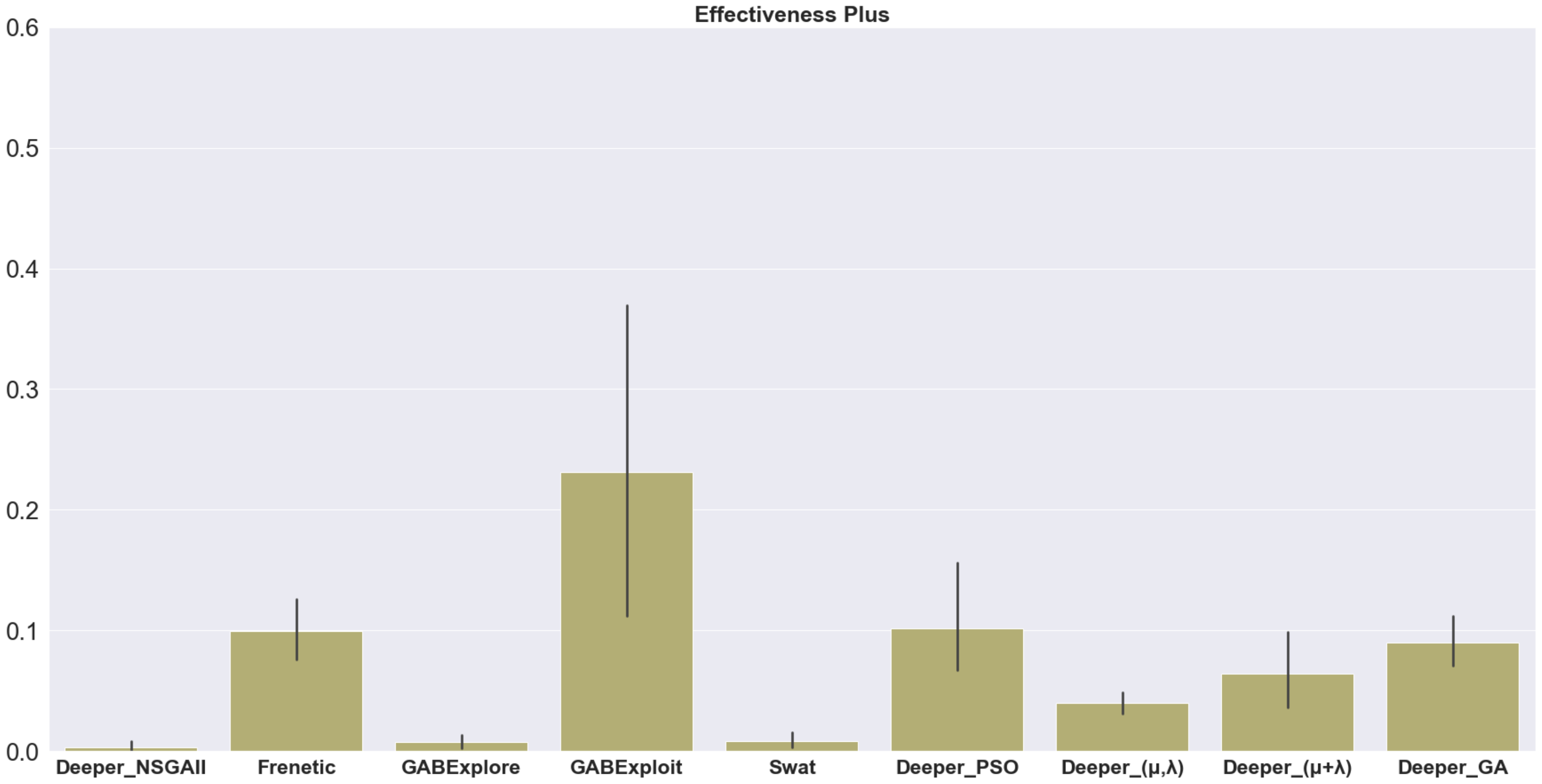}
         \caption{SET2}
         \label{fig:Exp_2_Effectiveness_Plus}
     \end{subfigure}
\caption{Test generation effectiveness plus in SET1 and SET2}
 \label{fig:Effectiveness_Plus_SET1_SET2}
\end{figure}

Figures \ref{fig:Exp_1_Effectiveness_Plus} and \ref{fig:Exp_2_Effectiveness_Plus} report the test effectiveness plus for the test generators in SET1 and SET2 respectively. In SET1, {Deeper} $(\mu+\lambda)$ ES-driven test generator results in the highest target-based effectiveness, and then {Deeper} GA, PSO, GABExploit, and Frenetic are the next effective tools. \pink{In this regard, Deeper PSO and GA also show a very comparable behavior to GABExploit.} \blue{In SET2, GABExploit shows the highest effectiveness plus, while Frenetic, {Deeper} PSO and  GA---with comparable performance---are the next ones leading to the highest ratio of triggered failures to the number of valid test scenarios.}  

\begin{table}[h]
\caption{\pink{Mann-Whitney U test on the test generation effectiveness plus in SET1 and SET2}}
    \begin{subtable}[h]{0.45\textwidth}
        \centering
        \caption{SET1}
        \begin{tabular}{|p{2cm}|p{2cm}|p{0.5cm}|p{0.5cm}|p{1cm}|}
        \toprule
        \textbf{Group1}       & \textbf{Group2}        & \textbf{U1}    & \textbf{U2}   & \textbf{p-value} \\ \midrule
        Deeper\_PSO           & Frenetic              & 12.0 &13.0 &1.0  \\
        \CLLL Deeper\_PSO           & GABExplore            & 25.0  &0.0  &0.0079 \\
        Deeper\_PSO           & GABExploit            & 17.0  &8.0  &0.4206  \\
        \CLLL Deeper\_PSO           & Swat                  & 25.0  &0.0  &0.0079  \\
        \CLLL Deeper\_PSO           & Deeper\_NSGAII        & 25.0  &0.0  &0.0079 \\
        \CLLL Deeper\_PSO           & Deeper\_$(\mu+\lambda)$  & 0.0  &25.0 & 0.0079  \\
        \CLLL Deeper\_PSO           & Deeper\_$(\mu,\lambda)$ & 25.0  &0.0  &0.0079  \\
        Deeper\_PSO           & Deeper\_GA            & 9.0  &16.0  &0.5476 \\
        \CLLL Deeper\_$(\mu+\lambda)$  & Frenetic              & 25.0  &0.0  &0.0079  \\
        \CLLL Deeper\_$(\mu+\lambda)$  & GABExplore            & 25.0  &0.0  &0.0079  \\
        \CLLL Deeper\_$(\mu+\lambda)$  & GABExploit            & 25.0  &0.0  &0.0079  \\
        \CLLL Deeper\_$(\mu+\lambda)$  & Swat                  & 25.0  &0.0  &0.0079  \\
        \CLLL Deeper\_$(\mu+\lambda)$  & Deeper\_NSGAII        & 25.0  &0.0  &0.0079  \\
        \CLLL Deeper\_$(\mu+\lambda)$  & Deeper\_$(\mu,\lambda)$ & 25.0  &0.0  &0.0079  \\
        \CLLL Deeper\_$(\mu+\lambda)$  & Deeper\_GA            & 25.0  &0.0  &0.0079  \\
        \CLLL Deeper\_$(\mu,\lambda)$ & Frenetic              & 0.0  &25.0 & 0.0079  \\
        Deeper\_$(\mu,\lambda)$ & GABExplore            & 22.0  &3.0  &0.0556  \\
        Deeper\_$(\mu,\lambda)$ & GABExploit            & 7.0  &18.0  &0.3095  \\
        \CLLL Deeper\_$(\mu,\lambda)$ & Swat                  & 25.0  &0.0  &0.0079  \\
        Deeper\_$(\mu,\lambda)$ & Deeper\_NSGAII        & 22.0 &3.0  &0.0556  \\
        Deeper\_$(\mu,\lambda)$ & Deeper\_GA            & 5.0  &20.0 &0.1508  \\
        Deeper\_GA            & Frenetic              & 20.0  &5.0  &0.1508  \\
        \CLLL Deeper\_GA            & GABExplore            & 24.0  &1.0  &0.0159  \\
        Deeper\_GA            & GABExploit            & 15.0  &10.0  &0.6905     \\
        \CLLL Deeper\_GA            & Swat                  & 24.0  &1.0  &0.0159 \\
        \CLLL Deeper\_GA            & Deeper\_NSGAII        & 24.0  &1.0  &0.0159  \\
        Deeper\_NSGAII        & Frenetic              & 0.0 &25.0  &0.0079  \\
        Deeper\_NSGAII        & GABExplore            & 7.0  &18.0  &0.3095  \\
        Deeper\_NSGAII        & GABExploit            & 4.0  &21.0  &0.0952  \\
        Deeper\_NSGAII        & Swat                  & 6.0  &19.0  &0.2222  \\
        Frenetic              & GABExplore            & 25.0  &0.0  &0.0079  \\
        Frenetic              & GABExploit            & 15.0 &10.0 &0.6905     \\
        Frenetic              & Swat                  & 25.0  &0.0  &0.0079 \\
        GABExplore            & GABExploit            & 5.0  &20.0  &0.1508  \\
        GABExplore            & Swat                  & 14.0  &11.0  &0.8413 \\
        GABExploit            & Swat                  & 20.0 &5.0 &0.1508 \\ \bottomrule
        \end{tabular}
       \label{table:MannWhitney_effectiveness_SET1}
    \end{subtable}
    \hspace{1cm}
    \begin{subtable}[h]{0.45\textwidth}
        \centering
        \caption{SET2}
        \begin{tabular}{|p{2cm}|p{2cm}|p{0.5cm}|p{0.5cm}|p{1cm}|}
        \toprule
        \textbf{Group1}       & \textbf{Group2}       & \textbf{U1} & \textbf{U2} & \textbf{p-value} \\ \midrule
        Deeper\_PSO           & Frenetic              & 41.0        & 59.0        & 0.5288           \\
        \CLLL Deeper\_PSO           & GABExplore            & 100.0       & 0.0         & 0.0              \\
        Deeper\_PSO           & GABExploit            & 36.0        & 64.0        & 0.315            \\
        \CLLL Deeper\_PSO           & Swat                  & 100.0       & 0.0         & 0.0              \\
        \CLLL Deeper\_PSO           & Deeper\_NSGAII        & 100.0       & 0.0         & 0.0              \\
        Deeper\_PSO           & Deeper\_$(\mu+\lambda)$  & 74.0        & 26.0        & 0.0753           \\
        \CLLL Deeper\_PSO           & Deeper\_$(\mu,\lambda)$ & 94.5        & 5.5         & 0.0003           \\
        Deeper\_PSO           & Deeper\_GA            & 44.0        & 56.0        & 0.6842           \\
        Deeper\_$(\mu+\lambda)$  & Frenetic              & 24.0        & 76.0        & 0.0524           \\
        \CLLL Deeper\_$(\mu+\lambda)$  & GABExplore            & 99.0        & 1.0         & 0.0              \\
        Deeper\_$(\mu+\lambda)$  & GABExploit            & 27.0        & 73.0        & 0.0892           \\
        \CLLL Deeper\_$(\mu+\lambda)$  & Swat                  & 95.0        & 5.0         & 0.0002           \\
        \CLLL Deeper\_$(\mu+\lambda)$  & Deeper\_NSGAII        & 100.0       & 0.0         & 0.0              \\
        Deeper\_$(\mu+\lambda)$  & Deeper\_$(\mu,\lambda)$ & 59.5        & 40.5        & 0.5288           \\
        Deeper\_$(\mu+\lambda)$  & Deeper\_GA            & 25.5        & 74.5        & 0.0753           \\
        \CLLL Deeper\_$(\mu,\lambda)$ & Frenetic              & 6.0         & 94.0        & 0.0003           \\
        \CLLL Deeper\_$(\mu,\lambda)$ & GABExplore            & 96.0        & 4.0         & 0.0001           \\
        \CLLL Deeper\_$(\mu,\lambda)$ & GABExploit            & 23.0        & 77.0        & 0.0433           \\
        \CLLL Deeper\_$(\mu,\lambda)$ & Swat                  & 95.0        & 5.0         & 0.0002           \\
        \CLLL Deeper\_$(\mu,\lambda)$ & Deeper\_NSGAII        & 98.0        & 2.0         & 0.0              \\
        \CLLL Deeper\_$(\mu,\lambda)$ & Deeper\_GA            & 5.0         & 95.0        & 0.0002           \\
        Deeper\_GA            & Frenetic              & 44.5        & 55.5        & 0.7394           \\
        \CLLL Deeper\_GA            & GABExplore            & 100.0       & 0.0         & 0.0              \\
        Deeper\_GA            & GABExploit            & 35.0        & 65.0        & 0.2799           \\
        \CLLL Deeper\_GA            & Swat                  & 100.0       & 0.0         & 0.0              \\
        \CLLL Deeper\_GA            & Deeper\_NSGAII        & 100.0       & 0.0         & 0.0              \\
        Deeper\_NSGAII        & Frenetic              & 0.0         & 100.0       & 0.0              \\
        Deeper\_NSGAII        & GABExplore            & 35.5        & 64.5        & 0.315            \\
        Deeper\_NSGAII        & GABExploit            & 13.0        & 87.0        & 0.0039           \\
        Deeper\_NSGAII        & Swat                  & 33.0        & 67.0        & 0.2176           \\
        Frenetic              & GABExplore            & 100.0       & 0.0         & 0.0              \\
        Frenetic              & GABExploit            & 36.0        & 64.0        & 0.315            \\
        Frenetic              & Swat                  & 100.0       & 0.0         & 0.0              \\
        GABExplore            & GABExploit            & 15.0        & 85.0        & 0.0068           \\
        GABExplore            & Swat                  & 49.0        & 51.0        & 0.9705           \\
        GABExploit            & Swat                  & 84.0        & 16.0        & 0.0089          \\
         \bottomrule
        \end{tabular}
        \label{table:MannWhitney_effectiveness_SET2}
     \end{subtable}
     \label{table:MannWhitney_effectiveness}
\end{table}

\subsection{Discussion}
\blue{The conclusion from the experimental results for the number of detected failures presents the following remarks about the performance of the test generators: The similarity between 
Frenetic and Deeper GA in terms of the number of detected failures and also resulting effectiveness plus comes from the common use of GA in both test generators. However, they have used quite different representation models in the formulation of the problem. Frenetic uses curvature values to represent roads and augments the test generation with a model-based heuristic to promote a higher level of road diversity.}

\blue{Besides Frenetic and Deeper GA, GABExplore and GABExploit also use GA, as the evolutionary mechanism, for the generation of test scenarios. They use a parametric approach to represent the roads, i.e., the roads are presented as parametric curves, Bezier Curves, based on a set of control points. GABExploit uses a randomly generated seed population over the whole search, while GABExplore restarts the search with a new random seed population, once it finds a failure-revealing test scenario. Generally, GABExploit shows better performance in terms of the number of detected failures and also effectiveness plus compared to GABExplore. While, GABExplore could promote failure diversity, given enough test time budget, which can be expected due to renewing the seed population throughout the search process. In comparison to other tools, due to the common use of GA again, GABExploit shows almost similar behavior, w.r.t all aspects, to Deeper GA in non-strict experimental conditions (e.g., SET1). Though, in strict experimental configuration, it led to better performance, in terms of the number of triggered failures and effectiveness plus, 
which could be due to the specific representation model, and accordingly the customization of the genetic operations used in GABExploit.} 

\blue{Furthermore, excluding the impact of additional heuristics for diversity promotion, PSO as a swarm intelligence optimization algorithm in Deeper PSO also leads to fairly comparable performance to the GA-driven ones, in particular, Deeper GA, GABExploit---w.r.t different aspects---in this problem, though it still shows a more consistent behavior due to its specific continual evolution mechanism. The PSO-driven test generator never left an experiment without triggering any failure, while it was not valid for the GA-based tools, such as GABExploit. Deeper PSO shows quite similar behavior in terms of the number of detected failures and effectiveness plus to the GA-based tools, in particular, Deeper GA and Frenetic. However, due to the lack of an extra diversity-promoting heuristic, it did not result in triggering a very high level of failure diversity.}

\blue{Finally, regarding the role of evolutionary strategies, it is worth noting that due to the pivotal difference between GA and ES in the selection process (Section \ref{D:Sec::Search_Algorithms_ES}), and the impact of this difference within non-restricted time budget, {Deeper} $(\mu+\lambda)$ ES could outperform all other test tools in terms of the number of detected failures and effectiveness plus, once it gave enough time. While in a restricted test time budget it still works well and shows comparable performance to the PSO-based generator and most of the GA-based tools such as Frenetic and Deeper GA.}

\blue{At the end, in addition to the number of detected failures and effectiveness, and concerning the failure diversity, it is worth remarking that in order to promote a considerably high level of failure diversity, it is essential to augment the search process with an additional diversity-promoting heuristic, e.g., based on the representation model. Otherwise, given the use of the same representation model, almost all these swarm and evolutionary algorithms lead to comparable levels of failure diversity.}

\subsection{Threats to Validity}
The evaluation of {Deeper} comes with a set of threats to construct, internal, and external validity of the results.

\textbf{\textit{Construct validity:}} The choices of the fitness function---the distance of the car position from the center of the lane---and also the metrics used for calculating the sparseness and indicating the diversity of the test scenarios (weighted Levenshtein distance) in this study are domain-specific. However, we have based our choices on the sound metrics adopted by other research works in the literature \cite{panichella2021sbst, gambi2019asfault, riccio2020model}. \blue{Moreover, we also used the same algorithm settings, e.g., population size, crossover and mutation probabilities, in the proposed test generators to alleviate the potential non-comparability between the algorithms.}

\textbf{\textit{Internal validity:}} The randomized nature of the used bio-inspired algorithms could be a source of threats to the internal validity of the results. In this regard, we tried to follow the guidelines given by Arcuri and Briand \cite{arcuri2014hitchhiker} for the evaluation and analysis of the results, and mitigate this threat by reporting the distribution and statistics of the results, e.g., using Box plots to show the results, \pink{conducting non-parametric statistical tests, i.e., Kruskal-Wallis and pairwise Mann-Whitney U tests, on different parts of the results, i.e., detected failures, promoted diversity, and the resulting effectiveness. Furthermore, the limited number of runs for each tool in the experimental configurations---due to the long execution time and also being compliant with the applied experimental settings at the SBST (2021) tool competition---is another source of threat to the internal validity of the results. However, in order to mitigate the effects of this potential threat, we have applied non-parametric statistical tests, i.e., Kruskal-Wallis and Mann-Whitney U tests, which are appropriate for the situations in which the sizes of the data groups are small and no assumption about the underlying data distribution is available.} 

\textbf{\textit{External validity:}} The choice of the test subject system is a potential threat to the external validity of the results. However, the test system in this study is one of the commonly used systems in self-driving cars \cite{nguyen2022crisce, birchler2022cost, birchler2022single}, and furthermore different ML models with various quality levels (i.e., different accuracy levels) could be deployed within the BeamNG.AI agent \cite{riccio2020model}, and in this regard, the proposed test scenario generation techniques can still be used. Nonetheless, it still offers one type of DL-based systems in self-driving cars, and further studies are required to address the testing of other DL-driven systems, meanwhile, we also keep the tool open for extensions, for example, to support the execution of test scenarios in other simulators. 

\section{Related Work}\label{D:Sec::RelatedWork}

Test input data that can reveal failures in the test subject is the most common generated test artifact in the literature related to testing of automotive AI systems \cite{riccio2020testing}. Depending on the test level and the test subject, the inputs could be images, for instance, as used in DeepTest \cite{tian2018deeptest}, or test scenario configurations as used in \cite{ebadi2021efficient}. A brief overview of the most common techniques used to generate the test data is as follows:

\textbf{Input data mutation.} This type of mutation involves generating new inputs based on the transformation of the existing input data. For instance, DeepXplore \cite{pei2017deepxplore} uses such input transformations to find the inputs triggering different behaviors between similar autonomous driving DNN models, while also striving to increase the level of neuron coverage. Moreover, in many studies, those transformations are based on metamorphic relations. DeepTest \cite{tian2018deeptest} applies different transformations to a set of seed images with the aim of increasing neuron coverage and uses metamorphic relations to find the erroneous behaviors of different Udacity DNN models for self-driving cars. DeepRoad \cite{zhang2018deeproad} uses a GAN-based metamorphic testing technique to generate input images to test three autonomous driving Udacity DNN models. It defines the metamorphic relations such that the driving behavior in a new synthesized driving scene is expected to be consistent with the one in the corresponding original scene. 

\textbf{Test scenario manipulation.} Another major category of the methods to generate test input data is based on the manipulation and augmentation of the test scenarios. Most of the works in this category use search-based techniques to go through the search space of the scenarios to find the failure-revealing or collision-provoking test scenarios. In this regard, simulators as a form of digital twins have played a key role to generate and capture those critical failure-revealing test scenarios. Simulation-based testing can act as an effective complementary solution to field testing, since exhaustive field testing is expensive, meanwhile inefficient, and even dangerous, in some cases. 

Accordingly, various testing approaches relying on the simulators have been presented in the literature and in this regard, search-based techniques have been frequently used to address the generation of failure-revealing test scenarios. Abdessalem et al. utilize multi-objective search algorithms such as NSGA-II \cite{ben2016testing} along with surrogate models to find critical test scenarios with fewer simulations and then at less computation time for a pedestrian detection system. In a following study \cite{abdessalem2018testing}, they use MOSA \cite{panichella2015reformulating}---a many-objective optimization search algorithm--- along with objectives based on the branch coverage and some failure-based heuristics to detect undesired and failure-revealing feature interaction scenarios for integration testing in a self-driving car. Further, in another study \cite{abdessalem2018testinglearnable}, they leverage a combination of multi-objective optimization algorithms (NSGA-II) and decision tree classifier models---referred to as a learnable evolutionary algorithm---to guide the search-based process of generating critical test scenarios and also to refine a classification model that can characterize the failure-prone regions of the test input space for a pedestrian detection and emergency braking system.
Haq et al. \cite{haq2021automatic} use many-objective search algorithms to generate critical test data resulting in severe mispredictions for a facial key-points detection system in the automotive domain. Ebadi et al. \cite{ebadi2021efficient} benefit from GA along with a flexible data structure to model the test scenarios and a safety-based heuristic for defining the objective function to test the pedestrian detection and emergency braking system of the Baidu Apollo (an autonomous driving platform) within the SVL simulator.   

Regarding the impacts of the simulators in this area, Haq et al. \cite{haq2021can} provide a comparison between the results of testing DNN-based ADAS using online and offline testing. Their results clearly motivate an increased focus on online testing as it can identify faults that never would be detected in offline settings---whereas the opposite does not appear to be likely. Our current study responds to this call, and motivates our work on systems testing in simulated environments. With regard to a different perspective, Borg et al. \cite{borg2021digital} discuss the consistency between the test results obtained from running the same experiments based on two different simulators and investigate the reproducibility of the results in both simulators. When running the same testing campaign in PreScan and ESI Pro-SiVIC, the authors found notable differences in the test outputs related to revealed safety violations and the dynamics of cars and pedestrians. 

\blue{Furthermore, with regard to the use of BeamNG.AI as the test subject and BeamNG.Tech as the simulation environment in ML testing, in particular ADAS testing, Nguyen et al. \cite{nguyen2022crisce} propose a test generator that uses accident sketches and with the help of image processing identifies the road segments, lane markings and vehicle positions, and generates the corresponding test scenarios in BeamNG simulator. Birchler et al. \cite{birchler2022cost} develop a test scenario selection (SDC-Scissor) that uses ML techniques, i.e., logistic regression, random forest and naive Bayes, to train ML models to identify safe and unsafe test scenarios before their execution. In their study, they also use Lane keeping system in BeamNG.AI as the test subject and employ AsFault \cite{gambi2019asfault} as a GA-based test tool to generate the simulation-based test scenarios. In another study, they use single and multi-objective GA to create a test prioritization technique that can prioritize the test scenarios before their execution \cite{birchler2022single}. Similar to the previous studies, the lane keeping system in BeamNG is used as the test subject, the experimentation is done within BeamNG simulator and the scenarios are generated by using AsFault and the ML-based SDC-Scissor test scenario selection. Generally, in this regard, besides genetic algorithms, which are very commonly used in this area, utilizing evolution strategies and particle swarm optimization to generate failure-revealing test scenarios is a quite novel study.}    

\section{Conclusion}\label{D:sec::summary}
{Deeper} in its extended version utilizes a set of bio-inspired algorithms, genetic algorithm (GA), $(\mu + \lambda)$ and $(\mu , \lambda)$ evolution strategies (ES), and particle swarm optimization (PSO), to generate failure-revealing test scenarios for testing a DL-based lane-keeping system. The test subject is an AI agent in BeamNG.tech's driving simulator. The extended {Deeper} contains four new bio-inspired test generators that leverage a quality seed population and domain-specific crossover and mutation operations tailored for the presentation model used for modeling the test scenarios. Failures are defined as episodes where the ego car drives partially out of the lane w.r.t a certain tolerance threshold. In our empirical evaluation we focused to answer three main questions: first, how many failures the test generators can detect, second, how much diversity they can promote in the failure-revealing test scenarios, and third how effectively and efficiently they can perform, w.r.t different target failure severity (i.e., in terms of tolerance threshold), available test budget, and driving style constraints (e.g., speed limits). Our results show that the newly proposed test generators in {Deeper} present a considerable improvement on the previous version and they are able to act as effective and efficient test generators that provoke a considerable number of diverse failure-revealing test scenarios for testing an ML-driven lane-keeping system. They show considerable effectiveness in meeting the target, i.e., detecting diverse failures, with respect to different target failures intended and constraints imposed. In particular, they act as more reliable test generators than most of the counterpart tools for provoking diverse failures within a limited test budget and with respect to strict constraints.  

As some directions for future work, we plan to apply the proposed approaches to testing further types of ML-based lane-keeping systems, i.e., more industrial ones and also in other state-of-the-art simulation platforms. We also plan to extend the approaches by applying machine learning-based techniques such as reinforcement learning or Generative Adversarial Networks (GANs) for empowering the discovery of failure-revealing test scenarios.

\section*{Acknowledgment}
This work has been funded by Vinnova through the ITEA3 European IVVES (\url{https://itea3.org/project/ivves.html}) and H2020-ECSEL European AIDOaRT (\url{https://www.aidoart.eu/}) and InSecTT (\url{https://www.insectt.eu/}) projects. Furthermore, the project received partially financial support from the SMILE~III project financed by Vinnova, FFI, Fordonsstrategisk forskning och innovation under the grant number: 2019-05871.

\bibliographystyle{IEEEtran}
\bibliography{references}

\vspace{12pt}

\end{document}